\documentclass{optica-article}

\journal{opticajournal} 
\usepackage{xr-hyper}
\articletype{Research Article}
\usepackage{graphicx}%
\usepackage{braket}%
\usepackage{lineno}
\usepackage{color}
\newcommand{\new}[1]{{\color{black}{#1}}}

\begin{document}

\title{Enhancing Quantum Memories with Light-Matter Interference}

\author{Paul M. Burdekin\authormark{1,\dag, *}, Ilse Maillette de Buy Wenniger\authormark{1,\dag,*}, Steven Sagona-Stophel\authormark{1,2,\dag}, Jerzy Szuniewicz\authormark{2}, Aonan Zhang\authormark{1}, Sarah E. Thomas\authormark{1}, Ian A. Walmsley\authormark{1}}

\address{\authormark{1}Department of Physics, Imperial College London,  London, SW7 2BW, United Kingdom\\
\authormark{2}Okinawa Institute of Science and Technology, Okinawa, Japan\\
\authormark{3}Faculty of Physics, University of Warsaw, Pasteura 5, Warszawa, 02-093, Poland\\
\authormark{\dag} These authors contributed equally to this work.}
\email{\authormark{*}pmb18@ic.ac.uk, imaillet@ic.ac.uk} 

\begin{abstract}
Future optical quantum technologies, such as quantum networks, distributed quantum computing and sensing, demand efficient, broadband quantum memories. However, achieving high efficiency without introducing noise, reducing bandwidth, or limiting scalability remains a challenge. Here, we present a new approach to enhance quantum memory protocols by leveraging constructive light-matter interference\new{, leading to an increase in memory efficiency without increasing atomic density or laser intensity}. We implement this method in a Raman quantum memory in warm Cesium vapor, and achieve more than a three-fold improvement in total efficiency reaching $(34.3\pm8.4)\%$, while retaining GHz-bandwidth operation and low noise levels. Numerical simulations predict that this approach can boost efficiencies in systems limited by atomic density, such as cold atomic ensembles, from $65\%$ to beyond $96\%$, while in warm atomic vapors it could reduce the laser intensity needed to reach a given efficiency by over an order-of-magnitude, exceeding $95\%$ total efficiency. Furthermore, our method preserves the single-mode nature of the memory at high efficiencies. This new protocol is applicable to various memory architectures, paving the way toward scalable, efficient, low-noise, and high-bandwidth quantum memories.
\end{abstract}

\section{Introduction}
Optical quantum memories are essential for synchronizing probabilistic quantum processes~\cite{Heshami2016, Lvovsky2009, Nunn2013, Kaneda:17}, both for enhancing the rates of local quantum operations for quantum computation~\cite{scalableQuantumComputing3}, and enabling entanglement distribution across large-scale photonic networks~\cite{Briegel1998, Duan2001, Kimble2008}. To be viable in these applications, quantum memories must achieve near-unity efficiency and fidelity, operate at GHz-scale bandwidths, while offering simplicity in operation and scalability~\cite{Sangouard2011, Reim2010}. Additionally, memories that are single-mode enable applications such as coherent mode filtering, high-dimensional quantum information encoding, and quantum parameter estimation~\cite{Gao2019, Nunn2013, Mazelanik2022, Donohue2018, raymer_interference, Raymer_2020}.\\

While significant progress has been made across various physical systems using different memory protocols~\cite{Heshami2016}, no single approach has yet met all of these stringent criteria simultaneously. A major outstanding challenge for all protocols is to reach high efficiencies while maintaining low-noise operation. This challenge is even more pronounced for the storage and retrieval of broadband signals due to the need for very strong light-matter interactions over large bandwidths~\cite{Nunn2007, Gorshkov2007}. Cold alkali ensembles operated near resonance have demonstrated efficiencies of $92\%$ using electromagnetically induced transparency (EIT)~\cite{Hsiao2018} and $30\%$ with Autler-Townes Splitting (ATS)~\cite{Saglamyurek2021}. Here, the limited light-matter interaction due to low atomic densities (optical depth) necessitates on-resonance operation or the use of optical cavities, both of which restricts them to narrow bandwidths that are unsuited to many quantum light sources~\cite{Rakher2013, ThomasScienceAdv2024}.\\

Contrastingly, Raman-based memories~\cite{Kozhekin2000, Reim2010} in warm vapors allow for high optical depths and storage of broadband signals in the GHz-THz regime by leveraging off-resonant interactions. Demonstrations of Raman memories with warm atomic ensembles have reached efficiencies up to $82\%$~\cite{Guo2019}. However, the weaker light-matter interaction inherent to off-resonant schemes limits the efficiency and thus typically requires high control field energies~\cite{Guo2019, Thomas2019_FWM}, or the use of optical cavities~\cite{Saunders2016}. High control field energies increase noise, thus reducing the fidelity of the retrieved quantum state~\cite{Michelberger2015}, and also affect the single-mode nature of the memory, while cavities again constrain the acceptance bandwidth of the memory.\\

Here we introduce a novel approach to overcome these limitations and improve the performance of both resonant and off-resonant, cold and warm optical memory protocols: \textit{EEVI} - Efficiency Enhancement via light-matter Interference. Our protocol makes use of the beam splitter-like nature of an optically controlled quantum memory, and operates it in a split-step arrangement which can be conceptualized as a light-matter equivalent of a Mach-Zehnder interferometer. Interference between the optical field and matter-based excitation enables a phase-dependent enhancement of the memory efficiency without the need to increase the light-matter coupling or sacrificing memory bandwidth. We experimentally implement our method using a high-bandwidth Raman memory in warm Cesium vapor, achieving a more than three-fold improvement in total efficiency, reaching $(34.3\pm8.4)\%$ for an input signal bandwidth of $1~$GHz, without introducing additional noise. Furthermore, our numerical model predicts that EEVI can enable efficiencies approaching unity for high-bandwidth signals, while significantly reducing the need for large optical depths and intense control fields. This positions EEVI as a promising solution to the trade-offs between efficiency, bandwidth, and scalability that have long limited the practical application of optical quantum memories.\\

\section{The EEVI concept}
\begin{figure*}[t] 
\centering
\includegraphics[width=\textwidth]{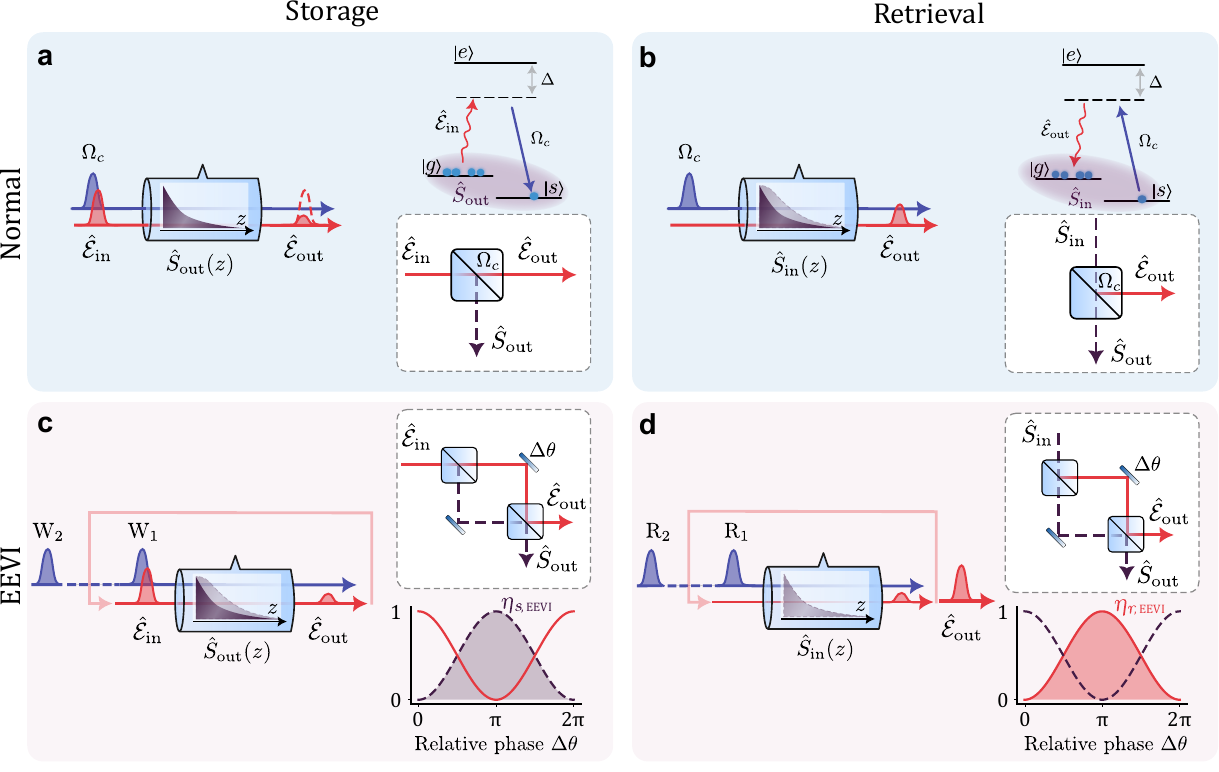}
\caption{The EEVI memory concept using the beam splitter analogy. \textbf{a.} The storage process of light using an ensemble-based Raman quantum memory. A strong write field $\Omega_c$ is temporally overlapped with an input signal field $\hat{\mathcal{E}}_\mathrm{in}$ at the memory, mediating the two-photon transition between hyperfine ground states $\ket{g}$ and $\ket{s}$ and generating a collective coherence called a spin wave $\hat{S}_\mathrm{out}$. The non-stored light is transmitted through the memory into output mode $\hat{\mathcal{E}}_\mathrm{out}$. In the single-mode regime, this interaction can be viewed as a beam splitter interaction where the input light couples to the modes  $\hat{S}_\mathrm{out}$ and $\hat{\mathcal{E}}_\mathrm{out}$ and the reflectivity (transmissivity) is determined by the control write field amplitude (inset).
\textbf{b.} After a certain storage time, a strong control read field is sent into the memory, reversing the process and thereby retrieving the stored light into output mode $\hat{\mathcal{E}}_\mathrm{out}$. In the beam splitter analogy (inset) this interaction couples an input spin wave $\hat{S}_\mathrm{in}$ to output modes $\hat{S}_\mathrm{out}$ and $\hat{\mathcal{E}}_\mathrm{out}$.
\textbf{c.} The EEVI procedure applied to the storage process. After initial storage with control write field W$_1$, the non-stored light is looped back into the memory input, overlapped with a second control write field W$_2$ and subsequently interferes with the previously created spin wave. In the beam splitter analogy (inset), this full process can be viewed as a Mach-Zehnder interferometer where an input signal field results in the interference of a spin wave and the non-stored light at the second beam splitter. The relative phase $\Delta\theta$ between the two interfering fields dictates the storage efficiency $\eta_{s,\mathrm{EEVI}}$, and for full constructive interference it can reach $100\%$ even when the memory is operating at $50\%$ efficiency. 
\textbf{d.} When EEVI is applied to the retrieval procedure, the light initially retrieved with control read field R$_1$ is sent back into the memory input, overlapped with a second control read field R$_2$, and interferes with the remaining spin wave, resulting in an enhanced retrieval efficiency $\eta_{r,\mathrm{EEVI}}$ for certain relative phases $\Delta\theta$. In this case, the input to the Mach-Zehnder is a spin wave.
}\label{fig1}
\end{figure*}

Optical quantum memory protocols fundamentally rely on light-matter interactions to coherently map an incoming optical field to a spatially-extended excitation across an ensemble of absorbers such as atoms. Often such a quantum memory is operated in a lambda configuration (Fig.~\ref{fig1}\textbf{a}), whereby an input photonic signal $\hat{\mathcal{E}}_\mathrm{in}(t)$ is temporally overlapped with a strong control field $\Omega_c(t)$ at the input of the memory, and subsequently stored as a coherence between the two ground states $\ket{g}$ and $\ket{s}$ -- referred to as a spin wave~\cite{Fleischhauer2002}. Following Ref.~\cite{Wasilewski2006}, for an adiabatic memory the coupling between the photonic field $\hat{\mathcal{E}}(t)$ and spin wave $\hat{\mathcal{S}}(z)$ input and output modes is written as:
\begin{subequations}
\begin{align}
    \hat{\mathcal{E}}_\mathrm{out}(t) &= \int_{0}^{T} dt' \, T_{\mathcal{E}\rightarrow\mathcal{E}}\left(\Omega_c(t),t, t'\right)\hat{\mathcal{E}}_\mathrm{in}(t') +  \int_{0}^{L} dz' \, R_{\mathcal{S}\rightarrow\mathcal{E}}\left(\Omega_c(t), z', t\right)\hat{\mathcal{S}}_\mathrm{in}(z') \\
    \hat{\mathcal{S}}_\mathrm{out}(z) &= \int_{0}^{L} dz' \, T_{\mathcal{S}\rightarrow\mathcal{S}} \left(\Omega_c(t), z, z'\right)\hat{\mathcal{S}}_\mathrm{in}(z') -  \int_{0}^{T} dt' \, R_{\mathcal{E}\rightarrow\mathcal{S}} \left(\Omega_c(t), z, t'\right)\hat{\mathcal{E}}_\mathrm{in}(t') \, ,
\end{align}    
\label{eq:Q_scattering_matrix}
\end{subequations}
where the memory interaction takes place during time $t'\in[0,T]$, and the absorbing ensemble has a spatial extent $z'\in[0,L]$. \new{The kernels in Eq.~\ref{eq:Q_scattering_matrix} are complex, with kernel $T_{\mathcal{E}\rightarrow\mathcal{E}}$ ($T_{\mathcal{S}\rightarrow\mathcal{S}}$) describing the relation between the input and transmitted signal field (input and output spin wave), and kernel $R_{\mathcal{E}\rightarrow\mathcal{S}}$ ($R_{\mathcal{S}\rightarrow\mathcal{E}}$) describing the time-to-space mapping in the memory storage (retrieval) process.} Since this mapping describes a unitary transformation, the kernels can be expressed as multiple, independent couplings between temporal modes, $\psi_{i}(t)$, and spatial modes, $\phi_{i}(z)$. We can use these modes to expand $\hat{\mathcal{E}}_\mathrm{in/out}(t)$ and $\hat{\mathcal{S}}_\mathrm{in/out}(z)$ in terms of their respective annihilation operators, $\hat{e}^{(i)}$ and $\hat{s}^{(i)}$, acting on modes $\psi_{i}(t)$ and $\phi_{i}(z)$ respectively (see Supplementary), and we find that Eq.~\ref{eq:Q_scattering_matrix} can be written in a form resembling the well-known beam splitter equation:
\begin{align}
\left(\begin{array}{c}
   \hat{e}_\mathrm{out}^{(i)}  \\
       \hat{s}_\mathrm{out}^{(i)} \\
\end{array}\right) = \left(\begin{array}{cc}
 t^{(i)} & r^{(i)}   \\
 -r^{(i)} & t^{(i)} \\
\end{array}\right) \left(\begin{array}{c}
   \hat{e}_\mathrm{in}^{(i)}  \\
       \hat{s}_\mathrm{in}^{(i)} \\
\end{array}\right)  \, ,
\label{eq:beamsplitter_simple}
\end{align}
where the transmissivity and reflection coefficients $t^{(i)}$ and $r^{(i)}$, as well as the temporal and spatial modes involved, $i$, are determined by the control field $\Omega_c(t)$~\cite{Nunn2008}. \new{Since the memory interaction is unitary, there is a minus sign in front of one of the reflection coefficients in Eq.~\ref{eq:beamsplitter_simple}~\cite{Wasilewski2006}.} In the low-coupling single-mode regime, only $r_{1}$ and $t_{1}$ are non-zero, and therefore determine the storage (retrieval) efficiency: $\eta_{s(r)} = |r_1|^2$. Thus, the memory interaction can be viewed as a beam splitter, where the input field is partially mapped onto an output spin wave mode and partially transmitted as an output optical field, see in Fig.~\ref{fig1}(\textbf{a,b}). An ideal quantum memory would have 100\% reflectivity for both the write and read interactions, i.e. perfect storage and retrieval. However, achieving high \new{efficiencies} ($>90\%$) for high-bandwidth input signals requires large optical depths and control field powers, as well as pulse shaping of the control pulses~\cite{Nunn2007, Guo2019, Gorshkov2007}. These requirements are often difficult to attain, lead to additional noise~\cite{Heshami2016}, nonlinear effects~\cite{Ara2013, Wu2022}, sacrifice the single-mode nature of the memory~\cite{Nunn2007} \new{and lead to incoherent losses}. Notably, achieving a larger efficiency becomes increasingly more challenging the closer it is to unity and, for example, increasing from 90\% to 99\% can require orders of magnitude higher control intensities (see Discussion). \\


Instead, we propose a new method to enhance the storage and retrieval efficiency of quantum memories by using light-matter interference. The general concept is shown in Fig.~\ref{fig1}(\textbf{c,d}), respectively, and can be conceptualized as a Mach-Zehnder (MZ) interferometer applied to either the storage or retrieval processes of the memory, or both. To apply EEVI to the storage interaction, a first write control pulse W$_1$ is chosen such that it couples $50\%$ of the input field $\hat{\mathcal{E}}_\mathrm{in}$ to the spin wave and $50\%$ of the signal is transmitted as a photonic field, analogous to a 50:50 beam splitter. The non-stored signal is looped back into the memory where a second write field W$_2$ (also with $50\%$ coupling) mediates interference between the spin wave and the looped photonic field, \new{see Supplementary. Depending on the relative phase $\Delta\theta$ between the photonic field and the spin wave, resulting from optical path changes or spin wave phase accumulation due to the presence of a magnetic field for example, the interference during the second pass can be constructive or destructive. This interference can also be varied by introducing a relative phase between the two control write fields.} For perfect mode overlap \new{and a certain phase difference}, this can lead to complete constructive interference, resulting in perfect storage efficiency $\eta_{s, \mathrm{EEVI}} =1$. This is equivalent to an all-optical MZ interferometer with two 50:50 beam splitters achieving full constructive interference into one output port. When considering loop loss, the storage efficiency after write field W$_2$ is given by
\begin{equation}
    \eta_{s, \mathrm{EEVI}} = \left\vert\frac{\mathcal{S}_\mathrm{out}}{\mathcal{E}_\mathrm{in}}\right\vert^{2} = \sqrt{\eta_{L}} \sin^{2}\left(\frac{\Delta \theta}{2}\right) + \frac{(1-\eta_L)^{2}}{4},
    \label{eq:storage_eff}
\end{equation}
where $\eta_L$ is the loop transmission, and $\Delta\theta$ is the relative phase between the looped signal and the spin wave.\\ 

We can also apply EEVI to the retrieval process, and can find an equivalent expression for the retrieval efficiency $\eta_{r,\mathrm{EEVI}}$. Here, the input to the light-matter interferometer is a spin wave $\hat{\mathcal{S}}_\mathrm{in}$, see Fig.~\ref{fig1}\textbf{d}. Upon applying a read field R$_1$, $50\%$ of the spin wave is mapped onto the signal field, and thus retrieved. The retrieved light is looped back to the input of the memory where another read field R$_2$ ($50\%$ mapping) enables light-matter interference. For full constructive interference, $100\%$ of the light can be retrieved from the memory. 

Finally, we can combine EEVI-storage and -retrieval, yielding a total efficiency proportional to $\eta_\mathrm{tot,EEVI}\propto\sin^4(\Delta\theta/2)$. \new{If each memory pass is set to $50\%$ efficiency (i.e. each storage and retrieval interaction individually achieves $50\%$), the full EEVI protocol can enhance the total efficiency to $100\%$ through constructive interference. In contrast, performing two independent storage and two independent retrieval steps -- without interference -- achieves at most $56\%$ total efficiency under the same conditions.} However, reaching perfect mode overlap to maximize the interference between the photonic and spin wave mode is nontrivial, but as shown later, numerical modeling demonstrates that with pulse shaping techniques a total memory efficiency approaching $100\%$ is possible.\\

This proposed method is similar to the all-optical Ramsey interferometer developed for the quantum pulse gate~\cite{raymer_interference}. It was shown through a two-step protocol, using the phase-dependence of the interaction in the second step, that high-efficiency mode filtering could occur. Our protocol makes use of the same split-step arrangement, but in contrast to interfering two light fields, we interfere a time-dependent photonic field with a spatially-extended matter excitation to increase memory efficiency. We note that this is distinct from the two-pass Raman memory in Ref.~\cite{Yu2023_doublepass} since it uses two independent memory interactions in opposite directions. \\


\begin{figure*}[h]
\centering
\includegraphics[width=\textwidth]{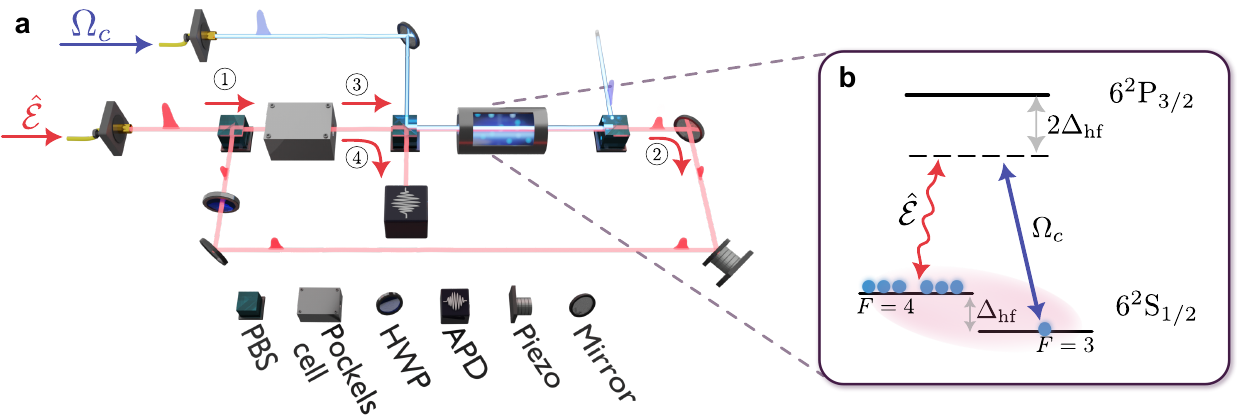}
\caption{The experimental implementation of EEVI. \textbf{a.} An input signal field $\hat{\mathcal{E}}$ (red) is sent into the setup \textcircled{1} and temporally overlapped at the memory with a strong control field $\Omega_c$ (blue) to map the signal field onto a spin wave. The non-stored signal field is looped back towards the input of the memory following path \textcircled{2} using polarization optics and a Pockels cell whereby the latter is triggered such that it either sends the light back to the memory \textcircled{3} or to the detectors \textcircled{4}. A piezo mounted to the back of a mirror inside the loop scans the optical phase, resulting in constructive or destructive interference between the spin wave and non-stored light when it is sent back into the memory \textcircled{3}. The same procedure can be applied to the retrieval process (see main). PBS: polarizing beam splitter; HWP: half-wave plate; APD: avalanche photodiode. 
\textbf{b.} The ensemble-based memory is operated in a Raman configuration and consists of Cesium-133 atoms initially optically pumped into the $\ket{F=4}$ hyperfine ground state. A strong control field drives the two-photon transition to the hyperfine ground state $\ket{F=3}$, creating a macroscopic coherence (spin wave) across the atomic ensemble between the two ground states. The signal and control field are both red-detuned from their respective transitions by $18.4$~GHz, which is twice the hyperfine ground state splitting $\Delta_\mathrm{hf}$.}\label{fig2}
\end{figure*}

\begin{figure*}[h]
\centering
\includegraphics[width=\textwidth]{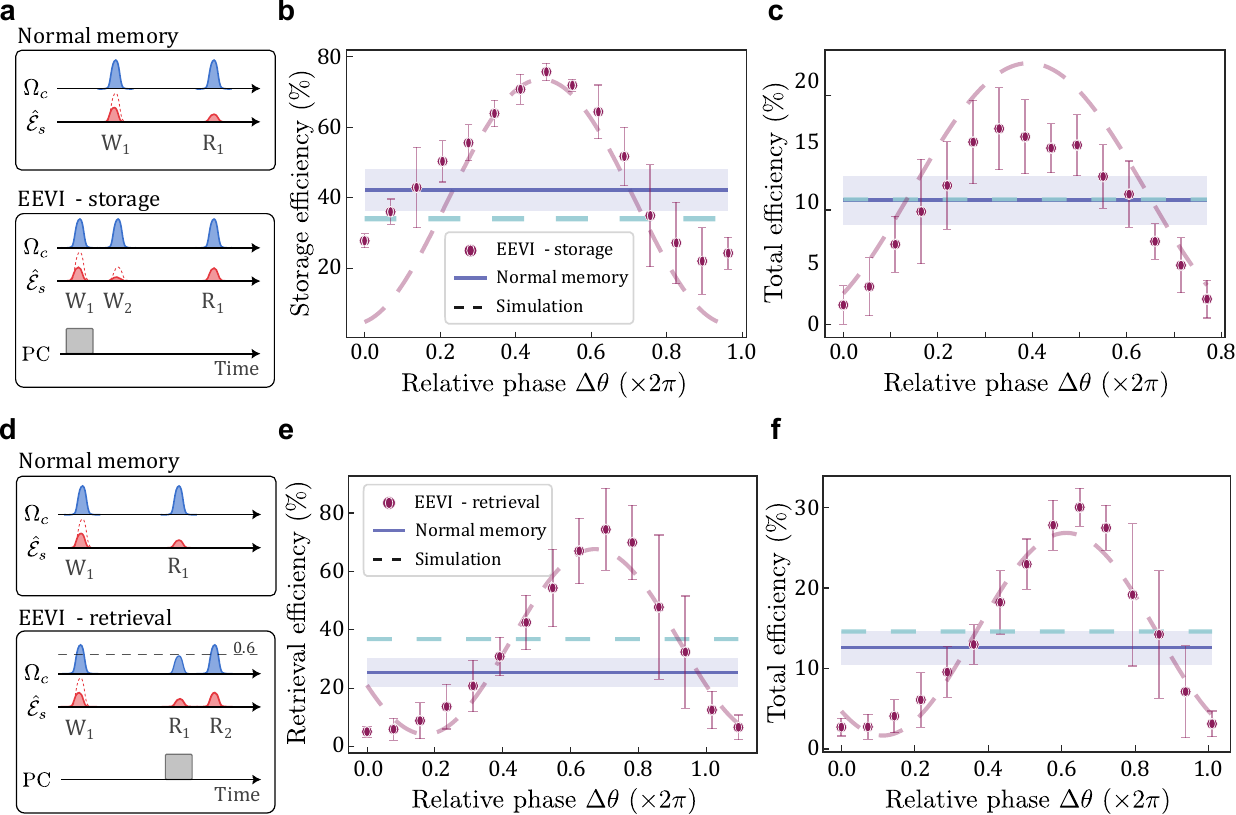}
\caption{Results and simulations for EEVI-Storage and -Retrieval. \textbf{a.} The pulse sequences for the write (W) and read (R) control $\Omega_c$ and signal $\hat{\mathcal{E}}_s$ fields used in normal Raman memory procedure (upper panel) and when applying EEVI to the storage process (lower panel). In the latter case a Pockels cell (PC) switches the non-stored light in bin W$_1$ back into the memory in bin W$_2$, where a second control write field enables interference between the looped signal field and previously stored spin wave.
\textbf{b.} The storage efficiency as a function of the relative optical phase $\Delta \theta$ for the EEVI-storage protocol (purple circles, where the error bars represent standard deviation across data sets). The storage efficiency for the equivalent normal Raman procedure is given by the solid blue line, and the shaded region indicates the standard deviation. The numerical simulations for EEVI-storage and the normal Raman memory are given by the purple and blue dashed lines, respectively. 
\textbf{c.} Same as \textbf{b} but for total efficiency. \textbf{d.} The pulse sequences used for EEVI-retrieval. \textbf{e.} The retrieval efficiency as a function relative optical phase $\Delta \theta$ for EEVI-retrieval and the equivalent normal memory. \textbf{f.} Same as \textbf{e} but for total efficiency.
}\label{fig34}
\end{figure*}

\section{Enhancing a GHz-bandwidth memory}

We experimentally demonstrate EEVI applied to a GHz-bandwidth Raman memory in a 7.5 cm Cesium vapor cell held at $82\,^{\circ}$C with 5 Torr of N$_{2}$ buffer gas. Initially, the atoms are prepared in the ground state $\ket{g}= \ket{6^2S_{1/2}, F=4}$ (fidelity $>99.7\%$, stationary optical depth of $5.3\times 10^{4}$) through optical pumping on the D1 transition.

We explore three scenarios: EEVI-storage, EEVI-retrieval, and EEVI-Raman (i.e. combining EEVI-storage and -retrieval) using the setup shown in Fig.~\ref{fig2}\textbf{a}. A signal field $\hat{\mathcal{E}}$ entering the memory setup through \textcircled{1} is temporally overlapped with a orthogonally polarized strong control field $\Omega_c$ at the memory, which drives the two-photon Raman transition shown in  Fig.~\ref{fig2}\textbf{b} thereby storing the signal field as a collective atomic ground state coherence. Both the signal and control field have a 330~ps duration, and are red-detuned by $18.4$~GHz from their respective transition to eliminate four-wave mixing~\cite{Thomas2019_FWM}. Applying a second control field reverses the process and retrieves the signal. After memory interaction the non-stored (or retrieved) light is sent into a \textcircled{2} free-space loop ($\eta_L = (63.5\pm2.5)\%$) containing a piezo. The light is redirected using a triggered Pockels cell (PC) which either sends the signal towards \textcircled{3} the memory \new{(with the same polarization as the original input signal)} or \textcircled{4} the detection setup. The piezo is used to scan the relative phase $\Delta\theta$ between the looped light and the spin wave (see Supplementary for details). \\

For EEVI-storage, the pulse sequence and PC timings are shown in Fig.~\ref{fig34}\textbf{a} together with the equivalent normal Raman memory which uses a single write (W$_1$) and read (R$_1$) pulse, and all control pulse energies are set to be the same: $400\,$pJ. The storage time, the time between the final write and first read control pulse, is $12.48$~ns in both cases. With these parameters, the normal Raman memory reaches an average storage efficiency of $(42.1\pm 5.0)\%$ and total internal memory efficiency of $(10.8\pm 2.1)\%$. For EEVI-storage, the PC switches the non-stored light after the first write process (bin W$_1$) back into the memory for a second write attempt (W$_2$), and $12.48$~ns later the spin wave is retrieved by a single read pulse R$_1$ (i.e. standard retrieval).\\

Fig.~\ref{fig34}\textbf{b} shows the storage efficiency of the EEVI-storage protocol as we scan $\Delta\theta$, as well as the average storage efficiency for the standard Raman memory (solid line). We observe a clear sinusoidal variation in storage efficiency when interfering the non-stored light with the spin wave. For constructive interference this leads to a two-fold improvement and a maximum storage efficiency of $(72.3\pm8.3)\%$, which is in agreement with Eq.~\ref{eq:storage_eff} for $\eta_L=(63.5\pm2.5)\%$. The dashed lines correspond to numerical simulations using experimentally measured parameters (Supplementary). This storage efficiency is comparable to state-of-the-art Raman memories~\cite{Guo2019}, but at six times larger detuning and 30-fold increase in bandwidth without any increase in control pulse energy. 

The more important metric, however, is the total efficiency. Indeed, an improvement in storage efficiency does not always correspond to an improvement in total efficiency, due to spin wave amplitude accumulation (bunching) at the input to the vapor cell, leading to re-absorption when retrieving in the forwards direction~\cite{Gorshkov2007, Nunn2007}. This effect is expected to be stronger in EEVI-storage, since any spin wave bunching after initial storage can lead to constructive interference concentrated at the input with the second write process. Regardless, we observe a sinusoidal trend in the total efficiency for EEVI-storage (Fig.~\ref{fig34}\textbf{c}), demonstrating again a two-fold increase in total efficiency to $(17.1\pm3.6)\%$. For storage efficiencies exceeding $72\%$, however, a lower total efficiency was observed partly due to spin wave bunching, but also due to nonlinear effects such as self-defocusing and self-phase modulation caused by high control field intensities (Supplementary). As we show later, these effects can be mitigated with EEVI through control pulse shaping. \\



Next, we apply EEVI to the retrieval procedure, but applying standard storage, using the timings shown in Fig.~\ref{fig34}\textbf{d}. Here the interference is between the spin wave and the initially \textit{retrieved} signal (R$_1$) switched into bin R$_2$. We set the control pulse energy to $800\,$pJ, resulting in a retrieval efficiency of $(25.3\pm5.0)\%$ and total efficiency of $(12.6\pm2.1)\%$ for standard Raman. For EEVI-retrieval, we again observe light-matter interference, but initially the observed maximum efficiency was reduced compared to EEVI-storage. This was primarily due to suboptimal overlap between the retrieved light in bin R$_1$ and remaining spin wave. We find that higher overlap is achieved when reducing the energy of the R$_1$ control pulse to $460\,$pJ. This leads to a greater interference visibility in the retrieval (Fig.\,\ref{fig34}\textbf{e}) and total efficiencies (Fig.\,\ref{fig34}\textbf{f}) as we scan $\Delta\theta$. With these pulse energies, we observe a more than two-fold increase in both efficiencies with respect to standard Raman, reaching a maximum retrieval efficiency of $(74.3\pm14.0)\%$ and total efficiency of $(30.0\pm2.3)\%$. \\

\begin{figure*}[h]
\centering
\includegraphics[width=\textwidth]{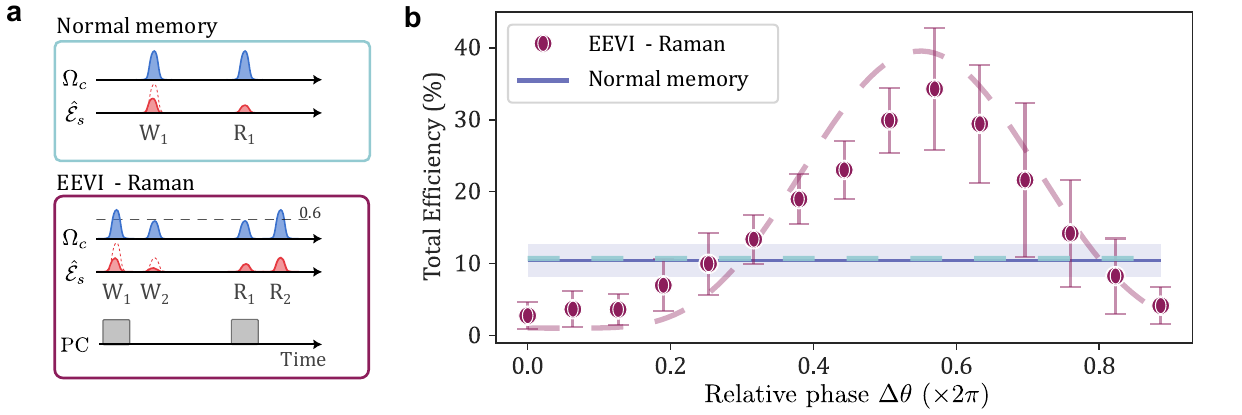}
\caption{Results and simulations for EEVI - Raman.  \textbf{a.} The pulse sequences for the write (W) and read (R) control $\Omega_c$ and signal $\hat{\mathcal{E}}_s$ fields used in normal memory procedure (upper panel) and when applying EEVI to the storage and retrieval procedure (lower panel). In the latter case a Pockels cell (PC) timing is set such that the transmitted light in bin W$_1$ and the retrieved light in bin R$_1$ are switched back into the memory (into bins W$_2$ and R$_2$, respectively), where the spin wave interferes with the light. The amplitudes of the control pulses used in the EEVI-Raman procedure are set to W$_2$/W$_1$ (R$_1$/R$_2$) $=0.6$ to optimize the efficiency. 
\textbf{b.} The total efficiency as a function of the optical phase $\Delta \theta$ for EEVI - Raman (purple circles). The total efficiency for the standard Raman memory is given by the blue solid line, and the shaded region indicates the standard deviation. The dashed lines are the numerical simulations. }\label{fig5}
\end{figure*}

For EEVI-Raman, applying EEVI to both storage and retrieval, we use the PC to switch the non-stored light in W$_1$ and the retrieved light in R$_1$ into the later time bins W$_2$ and R$_2$, respectively, see Fig.~\ref{fig5}\textbf{a}. Using a control pulse energy of $500$~pJ, we measure a total efficiency for the normal memory procedure of $(10.4\pm2.3)\%$. We optimize the total efficiency of EEVI-Raman by reducing the control pulse energy in bin W$_2$ and R$_1$ to $300$~pJ. As shown in Fig.~\ref{fig5}\textbf{b}, this combined procedure improves the efficiency over three-fold, reaching a total efficiency of $(34.3\pm8.4)\%$. In this case, the relative phase $\Delta\theta$ refers to both the write and read process. Conversely, the highest total efficiency we measure for the standard Raman memory is $(13.7\pm0.4)\%$ (Supplementary), obtained using control pulse energies of $1$~nJ, highlighting the increase in efficiency while reducing resources.  \\ 

\section{Noise performance}
Finally, we explore the noise performance of the EEVI-Raman memory. Noise in Raman memories can originate from four-wave mixing (FWM), fluorescence, spontaneous Raman scattering due to imperfect optical pumping, and insufficient filtering of the strong control field after the memory~\cite{Manz2007, Michelberger2015,Thomas2019_FWM}. In our demonstration we operate the memory at a specific detuning from resonance that offers built-in noise suppression by strongly absorbing the anti-Stokes field, thereby fully suppressing FWM noise~\cite{Thomas2019_FWM}. As such, we expect the noise in bin R$_1$ or R$_2$ to predominantly originate from the other three incoherent processes, which scale linearly with control pulse energy, are independent of the signal field and cannot build up or interfere between subsequent applications of the control pulse.

To quantify the remaining sources of noise, we perform the same measurements but with the signal field blocked. We observe no phase-dependence in the number of noise photons generated by the memory, and measure an average noise floor of $(2.2\pm1.1)\times10^{-2}$ photons per pulse for the EEVI-Raman protocol compared to $(3.9\pm0.2)\times10^{-2}$ for the equivalent normal Raman memory, demonstrating that the EEVI process does not increase the noise. For our demonstration using weak coherent states with an average input photon-number of $\bar{n}\sim 17$, this results in a signal-to-noise ratio (SNR) on retrieval of $47\pm20$ for standard Raman and $(187\pm104)$ for EEVI-Raman at $\Delta\theta$ corresponding to maximum constructive interference. The EEVI process thus increases the memory efficiency without increasing the noise, and therefore could enable higher fidelity storage and retrieval of quantum states. Since resonant memory protocols such as EIT and ATS are also dominated by linear processes such as fluorescence noise, we expect similar improvements in SNR when applying EEVI. 

\begin{figure*}[h]
\centering
\includegraphics[width=\textwidth]{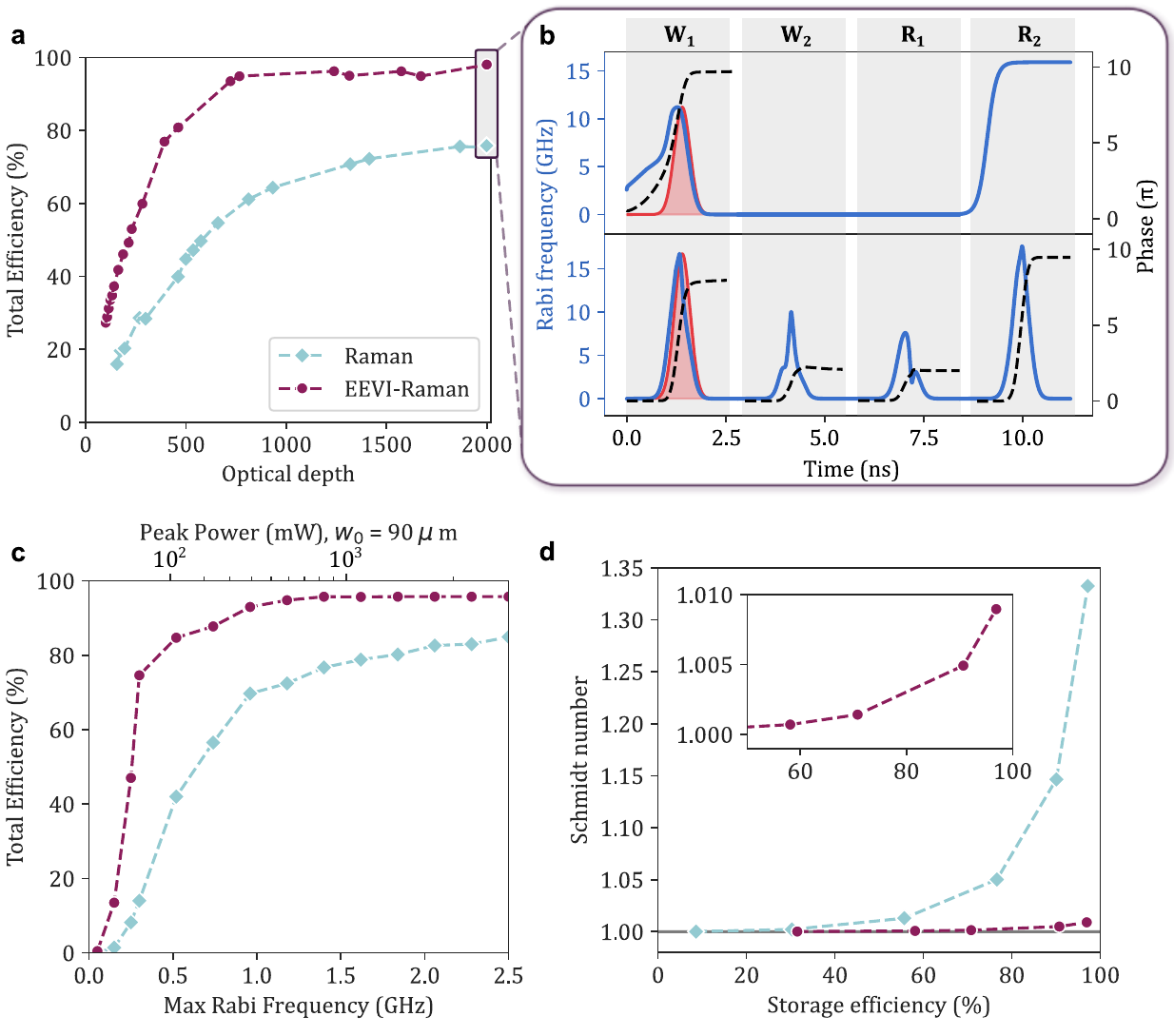}
\caption{\new{Simulation results using optimized} control pulses for Raman and EEVI-Raman. \textbf{a.} Total efficiency for forwards retrieval as a function of optical depth for the standard Raman memory (diamond) and EEVI-Raman (circle), assuming $100\%$ loop transmission \new{and for cryogenically cooled atoms.} \textbf{b.} An example of an optimized pulse sequence for an optical depth of 2000, for standard Raman (top) and EEVI-Raman (bottom), where the control field Rabi frequency and phase is plotted as solid blue and dashed black lines respectively, together with the input signal in red. \textbf{c.} Total efficiency for forward retrieval as a function of maximum peak control field Rabi frequency, for standard Raman memory (diamonds) and EEVI-Raman (circles), \new{where the temperature for the atoms is set to $80~^{\circ}$C}. Also shown on the top x-axis is the equivalent peak power, using the same parameters as the simulation of the experiment, with a focused beam waist of $w_{0}=90\,\mu$m. \textbf{d.} Schmidt number as a function of storage efficiency for standard Raman memory (diamonds) and EEVI-Raman (circles), where a Schmidt number of one indicates a fully single-mode memory. The inset shows a zoom in of the Schmidt number for EEVI-Raman as a function of storage efficiency.}\label{fig6}
\end{figure*}

\section{Discussion}
The experimental results of EEVI applied to a Raman memory show a clear improvement in memory efficiency, and the numerical simulations match the data well. Further improvements in efficiency can be made by optimizing the loop transmission and temporal shaping of the control field. The loop transmission can be readily improved through the use of low-loss optics -- a similar experimental setup with a free-space loop and Pockels cell has demonstrated a loop transmission of 98.8\%~\cite{Kaneda:17}. Our numerical model predicts that such an improvement to the loop transmission would result in a maximum total efficiency of $55\%$ with the same experimental parameters.\\

Temporal shaping of the control pulses can be implemented by using electro-optic modulators and optical amplifiers~\cite{Thomas2019_FWM}. Using our numerical model, and assuming $100\%$ loop transmission, we calculate the total efficiency for forward retrieval after a local optimization of the control field temporal profile. The numerical optimization follows the same procedure as laid out in Refs.~\cite{Gorshkov2008, Rakher2013} for complex control fields and retrieval in the forward direction (see Supplementary). We note that the efficiencies obtained here are not the theoretical maximum and that in principle, higher efficiencies could be reached through further optimization.\\

We explore the performance of EEVI in a Raman configuration in different parameter regimes, while keeping the signal bandwidth fixed at 1~GHz and with a red-detuning of $18.4~$GHz. In Fig.~\ref{fig6}\textbf{a} we investigate systems that are limited by optical depth such as cold atomic systems in magneto-optical traps (MOT) or micro-vapor cells. For instance, optical depths of 1000 have been demonstrated for MOTs~\cite{Tranter2018, Sparkes2013}, yet even after optimization the total efficiency in the forwards direction, the standard Raman protocol barely surpasses $60\%$. However, applying EEVI to the storage and retrieval processes (EEVI-Raman) improves the total efficiency to $>96\%$, without an increase in required control field intensity. Fig.~\ref{fig6}\textbf{b} shows the optimized control pulse sequences in blue for Raman (top) and EEVI-Raman (bottom) at an optical depth of 2000, where applying EEVI boosts the total efficiency from $75\%$ to $>98$\% without increasing the maximum Rabi frequency or changing the phase chirp significantly. As shown in the Supplementary, the same optimization procedure applied to resonant memories predicts similar improvements with EEVI.\\

For warm atomic vapors, the main limitation is often posed by the available control field intensity. For a Cesium ensemble kept at $80^{\circ}$C and restricting the maximum Rabi frequency, we show in Fig.~\ref{fig6}\textbf{c} the total efficiency obtained with optimized control pulses. Indeed, EEVI significantly reduces the control field Rabi frequencies needed and the improvements become more prominent at higher efficiencies where the total efficiency begins to plateau. For example, for a total efficiency of $80\%$ EEVI reduces the required peak Rabi frequency by more than four times, corresponding to a $>16$-fold reduction in peak intensity. We also highlight that a Rabi frequency of $1.4\,$GHz is sufficient to reach $>95\%$ total efficiency with EEVI, which for a beam waist of $90\,\mu$m as used in our experimental demonstration, corresponds to a peak power of $800\,$mW. This reduction in required control intensity not only lowers the technical requirements for efficient memories, but also reduces noise contributions and unwanted nonlinear effects (see Supplementary). For memories where FWM is present, we expect EEVI to show a significant reduction in noise since FWM scales quadratically with control field energy. EEVI thus enables a route towards efficient high-fidelity storage and retrieval of quantum states.\\

Lastly, we investigate the single-mode nature of EEVI applied to the \textit{storage} process and compare it to the equivalent standard Raman memory. The Raman memory has been shown to be single-mode in the low coupling regime~\cite{Nunn2008}, which means it can act as a coherent temporal mode filter~\cite{Gao2019}, but there is a trade-off between single-mode capacity and efficiency. Indeed, for a normal Raman memory the single-mode capacity of the storage process drastically decreases as a function of optimized storage efficiency, as shown in Fig.~\ref{fig6}\textbf{d} where we plot the simulated Schmidt number quantifying the modal capacity. We can see that for all storage efficiencies EEVI outperforms the standard Raman memory. This is expected since the EEVI protocol uses two subsequent memory interactions with a lower coupling strength and therefore each individual interaction remains more single mode, akin to the Ramsey interferometer for the quantum pulse gate~\cite{raymer_interference}. The most significant improvement is seen close to $100\%$ storage efficiency, where the Schmidt number is reduced from $1.33$ to $1.01$. EEVI could thus play a crucial role in technologies such as quantum parameter estimation and high-dimensional encoding, which require single-mode capacity and high efficiency. \\

In conclusion, we have demonstrated a new method to enhance the efficiency of existing optically-controlled quantum memory protocols, using quantum interference between an optical field and a stored material excitation. Experiments employing a high-bandwidth Raman memory show that applying this method results in a greater than three-fold improvement in memory efficiency without additional noise. Numerical modeling shows close agreement with our experimental results and predicts that with pulse shaping and improved setup transmission, near-unity efficiencies are realistically achievable, while retaining single-mode capacity and reducing requirements on the control field intensity and ensemble optical depth. These improvements to quantum memories are important in applications such as quantum networks, distributed quantum computing and sensing, and reducing the resource requirements is critical for real-world deployment. Besides the advantage of reducing technical overhead, we expect the EEVI method to reduce common noise processes such as four-wave mixing, spontaneous Raman scattering and control field leakage. The EEVI method thus provides a potential route towards scalable, efficient, low noise, high-bandwidth quantum memories.

\begin{backmatter}

\bmsection{Acknowledgment}
This work was supported by the Engineering and Physical Sciences Research Council (EPSRC) through the Distributed Quantum Computing and Applications grant (EP/W032643/1). J.S. acknowledges funding from the National Science Center (Poland) under grant number: 2022/45/N/ST2/04249. S.E.T. acknowledges an Imperial College Research Fellowship. The authors are grateful to M. Karpi\'{n}ski for discussions and comments, and to R. C. Schofield and for his comments on the manuscript. 

\bmsection{Disclosures} A patent has been filed (application number 2408438.6) by Imperial College Innovations Limited on this work, with all authors listed as inventors. I.A.W. is the Chairman and co-founder of ORCA Computing Ltd. 

\bmsection{Data Availability Statement}
Data underlying the results presented in this paper are not
publicly available at this time but can be obtained from the authors upon reason-
able request.

\bmsection{Supplemental document}
See Supplement 1 for supporting content.

\end{backmatter}
\newpage
\title{Enhancing Quantum Memories with Light-Matter Interference: supplemental document}

\section*{The beam splitter analogy}
To arrive to Eq.~2 for an optically controlled quantum memory, we follow the model laid out in Ref.~\cite{Wasilewski2006}.

For an optically controlled adiabatic memory the coupling between the photonic field $\hat{\mathcal{E}}(t)$ and spin wave $\hat{\mathcal{S}}(z)$ input and output modes is written as:
\begin{subequations}
\begin{align}
    \hat{\mathcal{E}}_\mathrm{out}(t) &= \int_{0}^{T} dt' G\left[\Omega_c(t),t, t'\right]\hat{\mathcal{E}}_\mathrm{in}(t') +  \int_{0}^{L} dt K_{R}\left[\Omega_c(t), z', t\right]\hat{\mathcal{S}}_\mathrm{in}(z') \\
    \hat{\mathcal{S}}_\mathrm{out}(z) &= \int_{0}^{L} dz' M\left[\Omega_c(t), z, z'\right]\hat{\mathcal{S}}_\mathrm{in}(z') -  \int_{0}^{T} dt' K_{S}\left[\Omega_c(t), z, t'\right]\hat{\mathcal{E}}_\mathrm{in}(t') \, ,
\end{align}    
\label{eq:Q_scattering_matrix_suppl}
\end{subequations}
where the memory interaction takes place during time $t'\in[0, T]$, and the absorbing ensemble has a spatial extent $z'\in[0,L]$. We can use the Bloch-Messiah rule to rewrite the kernels $G, M, K_\mathrm{S}$ and $K_\mathrm{R}$, describing the coupling between the light fields and spin wave fields, in terms of their common singular vectors and values: 
\begin{subequations}
\begin{align}
    G\left(t, t'\right) &= \sum_i t^{(i)} \Psi^{(i) *}_\mathrm{out}(t)\Psi^{(i)}_\mathrm{in}(t')\\
    M\left(z, z'\right) &= \sum_i t^{(i)} \Phi^{(i) *}_\mathrm{out}(z)\Phi^{(i)}_\mathrm{in}(z')\\
    K_{R}\left(z', t\right) &= \sum_i r^{(i)} \Psi^{(i) *}_\mathrm{out}(t)\Phi^{(i)}_\mathrm{in}(z')\\
    K_{S}\left(z, t'\right) &=\sum_i r^{(i)} \Phi^{(i) *}_\mathrm{out}(z)\Psi^{(i)}_\mathrm{in}(t'),
\end{align}    
\end{subequations}
where we removed the explicit dependence on $\Omega_c$ for readability, and $\Psi^{(i)}_\mathrm{in/out}(t)$ and $\Phi^{(i)}_\mathrm{in/out}(z)$ form a set of orthonormal bases for the input, output photonic field, and spin wave, respectively. We can understand this mapping as multiple independent beam splitter interactions, with $r_{i}$ and $t_{i}$ denoting the reflectivity and transmissivity for mode $i$, set by the temporal shape of the control field $\Omega_c$~\cite{Nunn2008}.

The functions $\Psi^{(i)}_\mathrm{in/out}(t)$ and $\Phi^{(i)}_\mathrm{in/out}(z)$ can be used to decompose the annihilation operators $\hat{\mathcal{E}}_\mathrm{in/out}(t)$ and $\hat{\mathcal{S}}_\mathrm{in/out}(z)$, resulting in:
\begin{subequations}
    \begin{align}
        \hat{\mathcal{E}}_\mathrm{in}(t) &= \sum_i \hat{e}^{(i)}_\mathrm{in} \Psi^{(i)*}_\mathrm{in}(t)\\
        \hat{\mathcal{E}}_\mathrm{out}(t) &= \sum_i \hat{e}^{(i)}_\mathrm{out} \Psi^{(i)*}_\mathrm{out}(t)\\
        \hat{\mathcal{S}}_\mathrm{in}(z) &= \sum_i \hat{s}^{(i)}_\mathrm{in} \Phi^{(i)*}_\mathrm{in}(z)\\
        \hat{\mathcal{S}}_\mathrm{out}(t) &= \sum_i \hat{s}^{(i)}_\mathrm{out} \Phi^{(i)*}_\mathrm{out}(z),
    \end{align}
\end{subequations}
where $\hat{e}^{(i)}_\mathrm{in/out}$ and $\hat{s}^{(i)}_\mathrm{in/out}$ are the annihilation operators acting on the input and output modes and undergo transformations analogous to a beam splitter, where requiring the transformation to be unitary results in the minus sign in Eq.~2~\cite{Wasilewski2006}.

\section*{Interference in an ensemble-based memory}

\begin{figure*}[t]
\centering
\includegraphics[width=\textwidth]{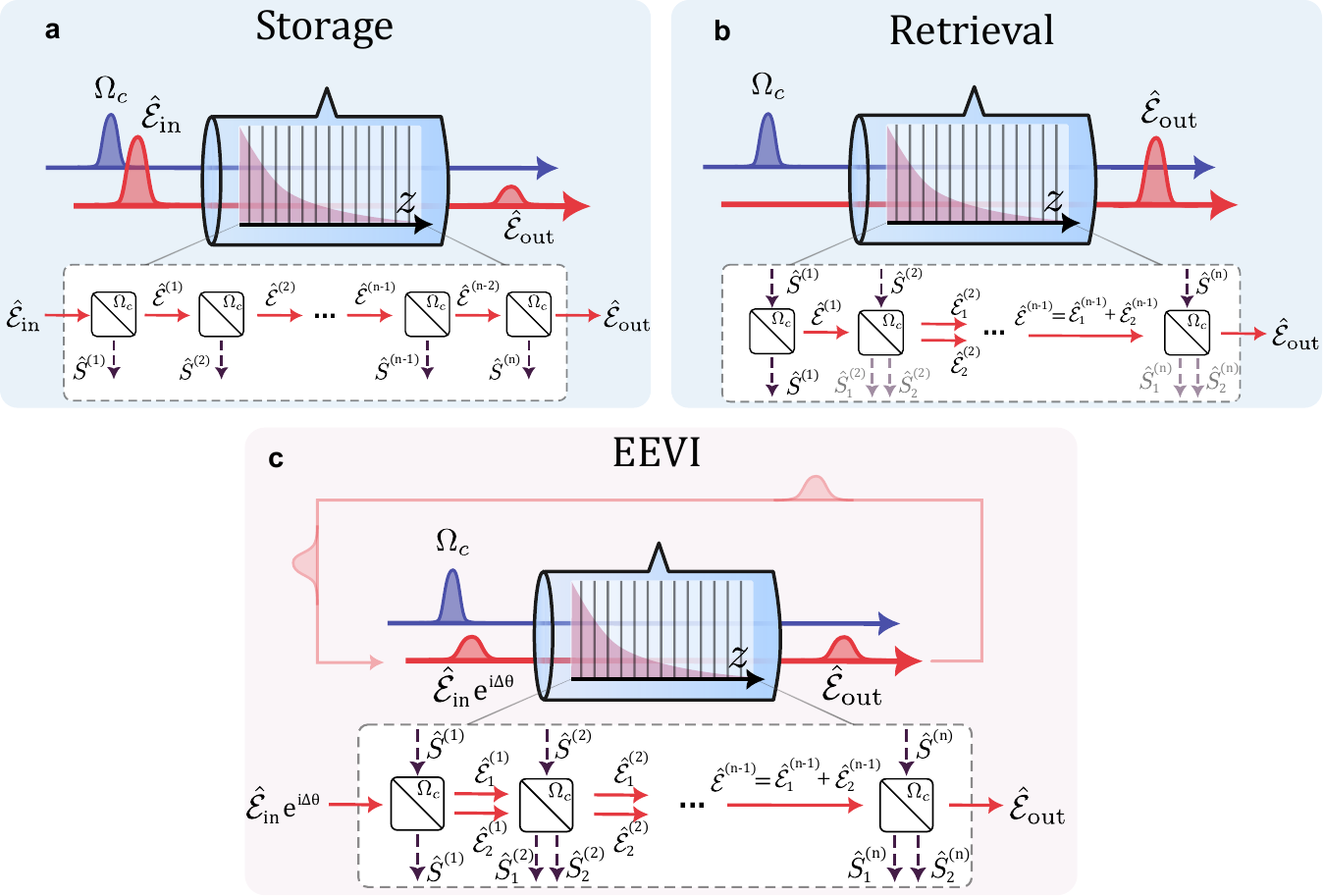}
\caption{An illustration of interference in an ensemble-based quantum memory. If we consider the ensemble to be made out of thin slices of atomic vapor along the propagation axis $z$, the memory can be viewed as a cascade of beam splitters. \textbf{a.} For a standard storage procedure with an input optical field, $\hat{\mathcal{E}}(z=0, t) = \hat{\mathcal{E}}_\mathrm{in}(t)$, the control field enables reflection (transmission) into the spin wave mode $\hat{S}^{(n)}$ (optical field $\hat{\mathcal{E}}^{(n)}$) at each slice $n$. \textbf{b.} For the retrieval process with an initial spin wave $\hat{S}(z, t=0)=\hat{S}_\mathrm{in}(z)$, a series of beam splitter interactions results in interference. Because the spin wave and the optical field are intrinsically phase-matched, the interference is constructive and results in re-emission of the stored light into output mode $\hat{\mathcal{E}}_\mathrm{out}$. \textbf{c.} For the EEVI protocol, after initial storage (retrieval), the non-stored (retrieved) light is looped back to the input of the memory for a second pass, where the light now acquires an optical phase $\Delta \theta$. Depending on the relative phase between this input field and the residual spin wave, this can result in enhanced storage (retrieval) efficiencies. See text for more details.}\label{figS}
\end{figure*}

Let us consider an ensemble-based quantum memory as depicted in Fig.~\ref{figS}. As laid out previously, such a memory acts as a time-varying beam splitter and by looping light after an initial storage or retrieval attempt back into the input of the memory, we can induce constructive or destructive light-matter interference between the looped light and the present spin wave depending on the relative phase $\Delta \theta$. In this section we give a physical interpretation of this interference and why the relative phase becomes critical when performing EEVI.\\

We first consider the standard memory storage process, see Fig.~\ref{figS}a. The memory cell, containing the atomic ensemble, is divided into $n$ infinitesimally thin slices along the light propagation axis $z$. Each slice is treated as a time-dependent beam splitter following the previous section, and as such, the overall interaction can thus be modeled as a cascade of beam splitter operations distributed along $z$. 

For the storage process, the boundary conditions are $\hat{S}_\mathrm{in}(z, t=0)=0$, i.e. no initial spin wave, and $\hat{\mathcal{E}}(z=0, t) = \hat{\mathcal{E}}_\mathrm{in}(t)$, representing the incoming signal field at the input of the memory. In the presence of a control field $\Omega_c$, the signal field undergoes a beam splitter-like transformation at each slice $n$: it is transmitted into the optical output field $\hat{\mathcal{E}}^{(n)}$, or reflected into the spin wave mode $\hat{S}^{(n)}$ with efficiencies $T$ and $R$, respectively, set by the control field. 

After the first slice, the transmitted component $\hat{\mathcal{E}}^{(1)}$ propagates towards the next slice where it undergoes another beam splitter interaction and the process repeats along the ensemble. In parallel, the reflected component $\hat{S}^{(1)}$ acquires a phase $\phi_{gs}$, which depends on the signal and control fields as well as the atomic dispersion. However, since only the forward-propagating optical field acts as an input to each slice, and the spin wave is stationary, there is no interference. Consequently, any phase carried by either the optical or spin wave field at a given slice $n$ has no observable effect on the storage process.\\ 

In contrast, the retrieval process is directly a consequence of interference within the memory resulting in the emission of light in the propagation direction~\cite{Gorshkov2007}. Fig.~\ref{figS}b conceptually depicts the retrieval process in the context of cascaded beam splitter interactions. The boundary conditions for the retrieval process are $\hat{S}(z, t=0)=\hat{S}_\mathrm{in}(z)$ and $\hat{\mathcal{E}}_\mathrm{in}(z = 0, t) =0$. Hence, at the first slice $n=1$ the input is solely the stationary spin wave $\hat{S}^{(1)}$ which is either reflected onto an optical field $\hat{\mathcal{E}}^{(1)}$, with a phase determined by the spin wave phase $\phi_{gs}$ and the retrieval control pulse, or transmitted into the same mode $\hat{S}^{(1)}$, the residual spin wave -- depending on the retrieval control field $\Omega_c$. 

Crucially, unlike the storage process, the input to slice $n>1$ during retrieval is not only the local spin wave $\hat{S}^{(n)}$, but also the optical field $\hat{\mathcal{E}}^{(n-1)}$ generated from the previous slice. These two fields combine and interfere at the beam splitter $n$. The spin wave phase is constant across the cell, up to a phase accumulated during storage due to dispersion. However, this same dispersion is present during retrieval, and so as the retrieved optical field propagates to the next slice, it will acquire the same phase shift due to dispersion, and therefore always be phase matched with the optical field retrieved from the spin wave at the current slice. The resulting interference is constructive and thus enables the mapping of the spin wave onto the forward propagating optical mode, visualized in Fig.~\ref{figS}b by the presence of two output modes constituting one optical mode $\hat{\mathcal{E}}^{(n)}=\hat{\mathcal{E}}^{(n)}_1+\hat{\mathcal{E}}^{(n)}_2$ (solid red arrows) and the resulting spin wave $\hat{S}^{(n)} =\hat{S}^{(n)}_1+\hat{S}^{(n)}_2$ (semi-transparent purple arrows), where the subscripts $1$ and $2$ denote the transmitted and reflected components respectively. As the process continues along the ensemble, the accumulated coherent interference between spin wave amplitudes and the optical field leads to directional re-emission of the initially stored light.\\

We now turn to the case of the EEVI protocol. The initial storage or retrieval process proceeds as discussed above. Afterwards, the non-stored (or retrieved) light is looped back to the input of the memory for a second interaction (potentially with different coefficients $R$ and $T$) and acquires an optical phase: $\hat{\mathcal{E}}_\mathrm{in}e^{i\Delta\theta}$, see Fig.~\ref{figS}c. At this second pass, the input to the first slice $n=1$ consists of both the looped optical field and the residual spin wave $\hat{S}^{(1)}$, thus resulting in interference which, importantly, can either be constructive or destructive depending on the relative phase $\Delta \theta$. By tuning this phase appropriately, interference throughout the cell can result in enhanced storage and retrieval efficiencies. This enables higher memory efficiencies for lower control field intensities compared to a standard memory procedure.

\section*{Experimental setup}

The full experimental setup is shown in Fig.~\ref{figS4}. Details on the setup are given below. In the experiments the laser at $351.7122$~THz, Pockels cells (PC), and pump EOM are all synchronized using an external clock. We collect photon time tags, piezo triggers, and PC triggers using a Swabian Instruments Time Tagger 20 time correlator and an avalanche photodiode. 

\begin{figure*}[h]
\centering
\includegraphics[width=0.9\textwidth]{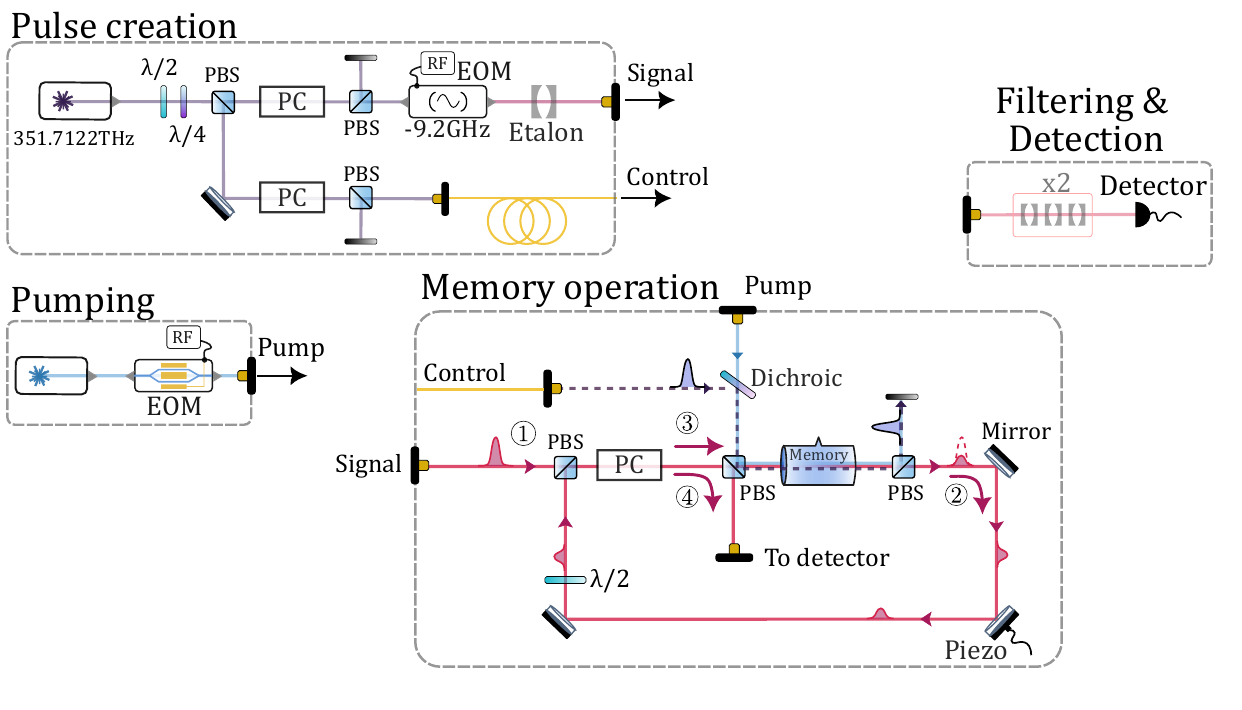}
\caption{The full experimental setup used during the experiments consists out of four parts: pumping, pulse creation, memory operation, and filtering and detection. Details are given in Methods.
}\label{figS4}
\end{figure*}

\subsection*{Memory operation}
The memory storage medium used is an atomic ensemble composed of Cesium (Cs-133) vapor held at $82\,^{\circ}$C within a $7.5\,$cm long vapor cell, together with 5 Torr of N$_{2}$ buffer gas. The cell is heated with quad-twisted cryogenic wire and is enclosed within a three-layer $\mu$-metal magnetic shield. We initially prepare the Cs atoms into the ground state $\ket{g}= \ket{6^2S_{1/2}, F=4}$ through optical pumping on the D1 Cesium transition using a continuous wave (CW) laser and an electro-optical modulator (EOM), see Fig.~\ref{figS4}. The EOM is triggered by an external clock and transmits the pump laser for approximately $2\mu$s, before being turned off during the memory operation. We obtain a pumping efficiency of  $>99.7\%$ of atoms in state $\ket{g}$. The ensemble has an on-resonant optical depth of $1240$, corresponding to a stationary optical depth of $5.3\times 10^{4}$.

We operate the memory in the off-resonant Raman configuration shown in Fig.~2\textbf{b}. We use the hyperfine ground states $\ket{6^2S_{1/2}, F=3}$ and $\ket{F=4}$, separated by $\Delta_\mathrm{hf}=9.2$~GHz, as our $\ket{g}$ and $\ket{s}$ states, respectively, and the excited state manifold $\ket{6^{2}\text{P}_{3/2}}$ as state $\ket{e}$ (i.e. operating on the D2 transition at $852~$nm).  
To operate the memory we apply a strong control field in two-photon resonance with a weak signal field, mediating the $\ket{g}\rightarrow\ket{s}$ transition. Both fields are red-detuned from the $\ket{e}$ state by $18.4$~GHz to eliminate four-wave mixing noise~\cite{Thomas2019_FWM}.

\subsection*{Pulse generation}
In our experiments, the timings of the control and signal pulse sequences, pump and Pockels cells are synchronized using an external clock. The control and signal pulse sequences are derived from a $330$~ps mode-locked Ti:Sapph laser tuned to the control frequency ($351.7122$~THz) and operating at a repetition rate of $80.13$~MHz. The pulses are split into two paths using a polarizing beam splitter, forming the weak signal field $\hat{\mathcal{E}}$ (average photon number $\bar{n}\approx 17$) and strong control field $\Omega_c$. Two Pockels cells, one in each arm, allow for pulse picking, after which the signal field is sent into an EOM generating sidebands at $\pm9.2$~GHz. A Fabry-P\'erot etalon with $36$~GHz free spectral range (FSR) at the output of the EOM filters out all but the red-shifted sideband, i.e. the signal frequency. Both the control and signal field are then sent into the memory setup where they are temporally overlapped using fiber delays.  A schematic of the pulse sequences used in the experiment is shown in Supplementary Fig.~\ref{figS2}.

\subsection*{Experiment}
The signal field enters the memory setup with a polarization aligned along the transmission axis \textcircled{1} of the polarizing beam splitter (PBS) shown in Fig.~2\textbf{a}, where it is temporally and spatially overlapped with the strong co-propagating orthogonally polarized control pulse. A lens in front of the memory (not shown) focuses both fields to a beam waist of $90\,\mu$m at the center of the cell. After memory interaction, the non-stored signal field is transmitted through the memory and subsequently sent back towards the input of the memory using a free-space loop with a total round-trip transmission of $(63.5\pm2.5)\%$ \textcircled{2}. A Pockels cell can switch the polarization of the non-stored light, and thus either sends the light to a detector \textcircled{4} (equivalent to a normal memory procedure) or back into the memory for EEVI-storage \textcircled{3}. In the latter case, a second write field is overlapped with the non-stored light at the memory, resulting in interference between the spin wave and the optical field. A piezo in the loop scans the relative phase $\Delta \theta$ between the spin wave and the looped signal field at a rate of 100~Hz resulting in constructive (destructive) interference translated into an enhanced (reduced) storage efficiency. After some variable storage time, a control read field converts the spin wave into an optical signal, which is either subsequently sent to the detectors (standard retrieval) or back into the memory (EEVI-retrieval) where it is overlapped with a second read field. Depending on the Pockels cell switching, we can thus perform a normal memory procedure, EEVI-storage, EEVI-retrieval, or EEVI-Raman (EEVI applied to both the storage and retrieval processes). More details on the pulse sequences and Pockels cell timings used can be found in the Supplementary.

When the light is sent towards the detector, an avalanche photodiode (APD, Excelitas SPCM-AQRH), the signal field is first spectrally filtered to remove any control field persisting after polarization filtering. The photonic field is sent through three monolithic etalons, passing each individual etalon twice. The first two etalons have a FSR of 18.4~GHz and the final etalon a FSR of 103~GHz. The total transmission through the filtering setup is approximately $18\%$. The resulting weak signal field is then detected by the APD and timestamps are recorded by a time correlator (Swabian Instruments Time Tagger 20). The correlator also collects the triggers produced by the external clock triggering the pulse sequence, and the triggers from the piezo, which are later used for data analysis.

\subsection*{Data acquisition and analysis}
For each measurement run we accumulate data over a period of 5~seconds. The acquired data after each run consists of a list of absolute photon arrival times, piezo triggers, and clock timings (triggering the signal and control field Pockels cells). We use the clock triggers to create a mask for data processing, allowing us to identify photon clicks from the different time bins in the pulse sequence, see Supplementary Fig.~\ref{figS2} for a schematic of the pulse timings.

Based on the timings of the piezo triggers we can assign a phase $\Delta \theta$ to individual photon time tags. We therefore bin the photon time tags in each piezo interval (10~ms) into 15 segments, where we assume the phase to be constant across each segment. Since the piezo is linearly scanning the relative phase $\Delta \theta \in[0, 2\pi]$, we can thus associate each of the time tags within a single piezo scan with a phase resolution of $2\pi/15$. To acquire enough photon statistics within each time bin, but not wash out the phase-dependence, we align and stack the time tags of ten piezo scans, forming one data set. We now have the number of clicks in each pulse bin, i.e.  W$_1, \mathrm{W}_2, \mathrm{R}_1, \mathrm{R}_2$, as a function of phase $\Delta \theta$. For each data set we fit a $\sin^2(\Delta \theta - \theta_0)$ to find the phase offset, $\theta_0$, caused by fluctuations in the loop path length over time, and align the data sets to account for this phase offset. We discard any data sets where the error in finding the phase offset is too high. Finally, we average over several data sets to account for experimental fluctuations and to obtain statistical uncertainties. The error bars shown for the data in Figures~3, 4 and 5 are the standard deviation across the data sets. 

To compare to the equivalent standard Raman memory without interference, we use the pulse sequences and energies given in the main text without Pockels cell switching. We accumulate and analyze the data in the same way as described above. We verify that for each data set we observe no variation in memory efficiency due to changes in the alignment when scanning the piezo. We then take the average and standard deviation of the efficiency across the entire data set, and these are the solid lines and shaded uncertainty regions given in Figures~3, 4 and 5. 

\subsection*{Detector saturation}
Since we are using a click detector (single-photon avalanche diode), but weak coherent states with $\bar{n}\approx0.2$ at the detector, we expect some of the pulses to contain more than one photon but only one detector click would be registered. Therefore the number of clicks is an underestimate of the actual number of photons. To account for this detector saturation effect, we correct the average number of measured photon clicks $n'$ according to~\cite{Aonan2019}:
\begin{equation}
    \bar{n} = -\ln\left(1-n'\right).
\end{equation}
 
\subsection*{Calculating efficiencies}
To calculate the efficiencies in our experiments, we add a reference pulse (REF) to the signal pulse sequence, see Supplementary Fig.~\ref{figS2}. This signal pulse does not overlap with a control pulse and therefore transmits straight through the memory.

For all memory protocols (EEVI and normal) we calculate the total efficiency by considering the retrieved light in the respective final retrieval bin, either R$_1$ or R$_2$ depending on the protocol. The total efficiency is then directly given by the ratio between the counts in the retrieval bin and the reference bin:
\begin{equation}
    \eta_\mathrm{tot} = \frac{N_\mathrm{R_i}}{N_\mathrm{REF}},
\end{equation}
with $i = 1$ or 2.

For the storage and retrieval efficiencies in the EEVI protocol we require knowledge of the loop transmission. Indeed, the number of photon clicks in detection bins W$_2$ and/or R$_2$ is affected by the loop losses. For each measurement run we therefore perform a loop transmission measurement whereby we block the control field and loop the signal field in bin W$_1$ into bin W$_2$ using the Pockels cell placed in the memory loop. The loop transmission is then given by the ratio between the reference (REF) pulse, which has undergone one loop, and the signal field in bin W$_2$, which has undergone two loops:
\begin{equation}
    \eta_\mathrm{loop} = \frac{N_\mathrm{W_2}}{N_\mathrm{REF}}.
\end{equation}

For EEVI-storage we calculate the storage efficiency (after two write control fields). We therefore also need to know the amount of non-stored light in bin W$_1$. This amount is directly given from the storage efficiency in the normal memory procedure $\eta_s$ where the non-stored light in bin W$_1$ is not looped: $\eta_\mathrm{nonstor} = 1-\eta_s$. The total storage efficiency for EEVI is then given by:
\begin{equation}
    \eta_{s, \mathrm{EEVI}} = 1 +\eta_\mathrm{nonstor}(\eta_\mathrm{loop} - 1)-N_\mathrm{W_2}/N_\mathrm{REF}
\end{equation}
where $N_\mathrm{W_2}/N_\mathrm{REF}$ is the percentage of non-stored light after the second write procedure. 

Finally, for EEVI-retrieval we find the retrieval efficiency by combining the previous results:
\begin{equation}
    \eta_{r, \mathrm{EEVI}} =\frac{\eta_\mathrm{tot}}{\eta_{s, \mathrm{EEVI}}},
\end{equation}
where $\eta_\mathrm{tot}$ is calculated using bin R$_2$. In the normal memory procedure the retrieval efficiency is given by $\eta_r=\eta_\mathrm{tot}/\eta_s$.

\section*{Pulse sequences and Pockels cell timings}
\begin{figure*}[h]
\centering
\includegraphics[width=0.8\textwidth]{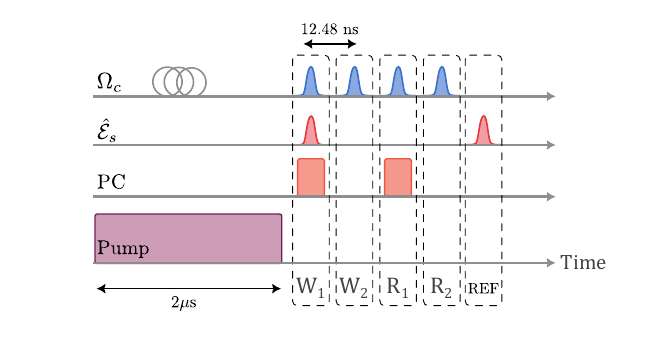}
\caption{The pulse sequences for the control field $\Omega_c$, signal field $\hat{\mathcal{E}}_s$, the Pockels cell (PC) in the EEVI loop, and the pump field. Depending on the measurement run, the PC switches the signal field into bin W$_2$ and/or R$_2$. The final signal pulse is used as a reference (bin REF).
}\label{figS2}
\end{figure*}

A schematic of the pulse sequences used in the experiments is shown in Fig.~\ref{figS2}. Depending on the experimental configuration (normal memory, EEVI-storage, EEVI-retrieval or EEVI-Raman), a Pockels cell in the memory loop (see Fig.~\ref{figS4}) switches the signal field in detection time bin W$_1$ (R$_1$) into the later detection time bin W$_2$ (R$_2$), meaning that the signal field is sent back into the memory (following path \textcircled{3}) where it interacts with another control field before being sent towards the detector (following path \textcircled{4}). We note that the signal pulse sequence contains also a reference pulse which is not overlapped with a control pulse at the memory. We use this reference pulse (bin REF) to calculate the memory efficiencies (see Methods).

\section*{Measuring the noise}
Raman memories can suffer from four-wave mixing (FWM) noise, which originates from the unwanted coupling of the control field to the populated ground state $\ket{g}$, in turn leading to spontaneous anti-Stokes scattering~\cite{Michelberger2015}. In our demonstration we operate the memory at a specific detuning from resonance that offers built-in noise suppression by strongly absorbing the anti-Stokes field, thereby fully suppressing FWM noise~\cite{Thomas2019_FWM}. In addition, Raman memories can suffer from other forms of noise such as fluorescence from the excited state $\ket{e}$ (if populated), spontaneous Raman scattering from any residual population in the storage state $\ket{s}$ due to imperfect optical pumping, and insufficient filtering of the strong control field from the weak signal field after the memory. For any of these three processes the noise scales linearly with the control pulse energy, and there is no coherent process that can build up or interfere between subsequent applications of the control pulse. We therefore expect that the noise in the retrieval time window R$_1$ or R$_2$ of the memory to be the same for the Raman and EEVI-Raman memory protocol if the control pulse energy in both cases is the same. \\

For each measurement in the main text (Figs.~3 and 4), we measure the number of noise photons per pulse $N_\mathrm{noise}$. This involves a complementary measurement whereby we block the signal field and measure the amount of light in bin R$_1$ or R$_2$, depending on the protocol. With the signal field blocked, the light in these bins can only result from the strong control field interacting with the atoms or being insufficiently filtered, i.e. the noise from the memory. The data is acquired and analyzed following the procedure described in the previous section. 

We calculate the noise using time bin R$_1$ for the normal Raman memory and EEVI-storage procedure, and R$_2$ for the EEVI-retrieval and EEVI-Raman procedure (see Figs. 3 and 4). As explained in the previous section, we accumulate the number of clicks over ten piezo scans and subsequently bin them into 15 phase bins, giving $t_\mathrm{int}$. The number of pulses within this integration window is then directly given by dividing $t_\mathrm{int}$ by the AWG trigger repetition rate $N_\mathrm{trigg}$. Hence we find that the number of noise photons per pulse in bin R$_1$ or R$_2$ is given by
\begin{equation*}
    N_\mathrm{noise}= \frac{N_\mathrm{clicks}}{t_\mathrm{int} N_\mathrm{trigg} \eta_\mathrm{transm} \eta_\mathrm{det}}, 
\end{equation*}
where $\eta_\mathrm{transm}$ and $\eta_\mathrm{det}$ are the transmission measured from the front of the cell to the detectors, and is $1.08\%$, and the detector efficiency, $\eta_\mathrm{det} =45\%$, respectively. The results for EEVI show no phase dependence in the noise. Since the noise photons per pulse in both the normal Raman memory and EEVI cases show no phase dependence, we can determine a signal-to-noise ratio (SNR) using the mean value of the noise. For all the measurements presented in the main, we find the average noise photons per pulse shown in Table~\ref{table1}, where the corresponding value for the normal Raman memory is given in square brackets. The SNR is calculated at maximum (total) memory efficiency, i.e. when we observe maximum constructive interference. 
\begin{table}[t]
\begin{tabular}{ |p{3.5cm}||p{3cm}|p{2.7cm}|p{2cm}|  }
 \hline
 \multicolumn{4}{|c|} {Noise measurement}\\
 \hline
 \textbf{Measurement \newline (relative pulse heights)}& \textbf{Noise per pulse \newline [ref. memory]} & \textbf{Total efficiency \newline [ref. memory]} & \textbf{SNR \newline [ref. memory]}\\
 \hline
 EEVI-storage  (1:1:1) &  $(2.6 \pm 0.9)\times 10^{-2}$  \newline $\left[(3.9 \pm 0.2)\times 10^{-2}\right]$  & $(17.1\pm3.6)\%$ \newline $\left[(10.8\pm2.1)\%\right]$ &  $121 \pm 45$ \newline $\left[47\pm20\right]$ \\
 \hline
 EEVI-retrieval (1:0.6:1) &  $(7.1 \pm 0.3)\times 10^{-2}$ \newline $\left[(14.7 \pm 3.2)\times 10^{-2}\right]$ &  $(30.0 \pm 2.3)\%$ \newline $\left[(12.6 \pm 2.1)\%\right]$ &$131\pm 55$ \newline $\left[25\pm6\right]$ \\
 \hline
EEVI-Raman (1:0.6:0.6:1) &  $(2.2 \pm 1.1)\times 10^{-2}$ \newline $\left[(3.9\pm 0.2)\times10^{-2}\right]$ & $(34.3\pm8.4)\%$ \newline $\left[(10.4\pm2.3)\%\right]$  &$187\pm 104$ \newline $[47 \pm 20]$\\
 \hline
\end{tabular}
\caption{The noise photons per pulse, maximum total efficiency (EEVI: for phase $\Delta \theta$ corresponding to maximum constructive interference; reference: average value) and signal-to-noise ratio (SNR) for the EEVI measurements presented in the main text, and the corresponding reference memory measurement (square brackets). \label{table1}}
\end{table}

\section*{Numerical simulations}
\subsection*{Solving Maxwell-Bloch equations}

The standard memory protocols presented in this work, together with their EEVI counterparts, were modeled by numerically solving the Maxwell-Bloch equations in Python, using a fourth-order Runge-Kutta method for solving the time dimension and a Chebyshev spectral method for solving the spatial dimension. We model the Cesium atom as a three-level system in a lambda configuration, where the ensemble as a whole is made up of many velocities $v$, distributed according to the 1-dimensional Maxwell-Boltzmann probability density function $f(v)$. For simulating the standard off-resonant Raman memory, where both the signal and control fields are set to a single-photon detuning of $\Delta=18.4\,$GHz (as was set in the experiment to eliminate four-wave mixing noise), the Maxwell-Bloch equations for the coupled photonic $\hat{\mathcal{E}}$ and atomic spin wave $\hat{\mathcal{S}}$ modes are:
\begin{subequations}
   \begin{align}
\partial_z \hat{\mathcal{E}}(z, \tau) &= \sum_{v} \Bigg( i\sqrt{d(v)} \frac{\Omega_c(\tau)}{\Gamma(v)} \hat{\mathcal{S}}(z, \tau, v) - \frac{d(v)}{\Gamma(v)} \hat{\mathcal{E}}(z, \tau) \Bigg)\\
\partial_\tau \hat{\mathcal{S}}(z, \tau, v) &= -i\Delta_{2}(v) - \frac{|\Omega_c(\tau)|^2}{\Gamma(v)}\hat{\mathcal{S}}(z, \tau, v) -i \sqrt{d(v)}\frac{\Omega^*_c(\tau)}{\Gamma(v)}\hat{\mathcal{E}}(z, \tau)  \, ,
\end{align} 
\label{eq:MB-Raman}
\end{subequations}
where $d(v) = d\, f(v)dv$ is the velocity-dependent optical depth, $\Omega_c(\tau)$ is the complex Rabi frequency of the control field, $\Gamma(v) = \gamma + \gamma_{c} + i \Delta - k_{s}v$ is the Doppler-shifted single-photon complex detuning, with $k_{s}$ the wavevector of the signal field and $\gamma_{c}$ the additional decoherence due to collisions with the $N_{2}$ buffer gas present in the Cs vapor cell. We also have the velocity-dependent two-photon detuning $\Delta_{2}(v) = (k_{c}-k_{s})v$, which for our configuration $k_{c}-k_{s}\approx 0$ and so can be neglected on the time scales considered in this experiment. When simulating a cold ensemble, all of the atoms are assumed to have zero velocity.\\

For the write process of a standard Raman memory, we set our boundary conditions to a Gaussian input signal and no initial spin wave, together with a control write pulse with complex amplitude. For the read process, we set the input photonic field to zero, and set the initial spin wave to the final spin wave of the write simulation. An input read control pulse converts some of the spin wave to an output photonic field, which we can integrate to compute the total efficiency of the memory.\\

When simulating EEVI-Raman, the first write process is equivalent to the standard Raman memory. For the second write process, the input boundary conditions are the output spin wave and photonic field from the first write simulation, but with the photonic field multiplied by a phase term to account for the phase accumulated while traveling around the loop, and a loss term to account for the finite loop transmission. We assume no phase has been accumulated by the spin wave, and that the two write control fields have no relative phase difference. Note that the loss term is the square root of the measured loop transmission as it is the photonic field \textit{amplitude} which is input into the simulation.  The same procedure is used for the second read process.

\subsection*{Fitting to experiment}

\paragraph{Optical depth}

We will now detail how the aforementioned numerical model was used to reconstruct the experimental data. We first calibrate the optical depth term ($d(v)$) by using the resonant Maxwell-Bloch equations for a inhomogeneously broadened three-level system (detailed in Ref.~\cite{Gorshkov2007b}) with no control field, and setting a continuous wave (CW) photonic field, normalized to contain only one photon over the simulation time. The time of the simulation was set to run until a steady state is reached ($\sim 100\,$ns). Any broadening due to the finite bandwidth of the simulation is assumed to be negligible. We include a pressure broadening of $100\,$MHz due to 5 Torr of N$_{2}$ buffer gas and step the detuning of the input photonic field with respect to the atomic transition. An experimentally measured spectra was fit using the ElecSus package~\cite{Zentile2015} to obtain the temperature and number density of the Cs atomic ensemble, which was combined with the reduced dipole moment of the $\ket{g}\rightarrow \ket{e = 6^{2}\text{P}_{3/2}}$ transition~\cite{Steck2024} to yield an optical depth. Comparing the resulting transmission from this numerical simulation to the experimentally measured spectra, we find a discrepancy of $1/2.2$, which we attribute to the three-level system approximation ($\sim 1/3$ for averaging over the dipole operator, combined with neglecting the hyperfine manifold spacing). Note that the optical depth defined in Eqs.~\ref{eq:MB-Raman} is $1/2$ of the standard definition and this factor has already been taken into account.\\

\paragraph{Control field Rabi frequency}

\begin{figure*}[t]
\centering
\includegraphics[width=0.8\textwidth]{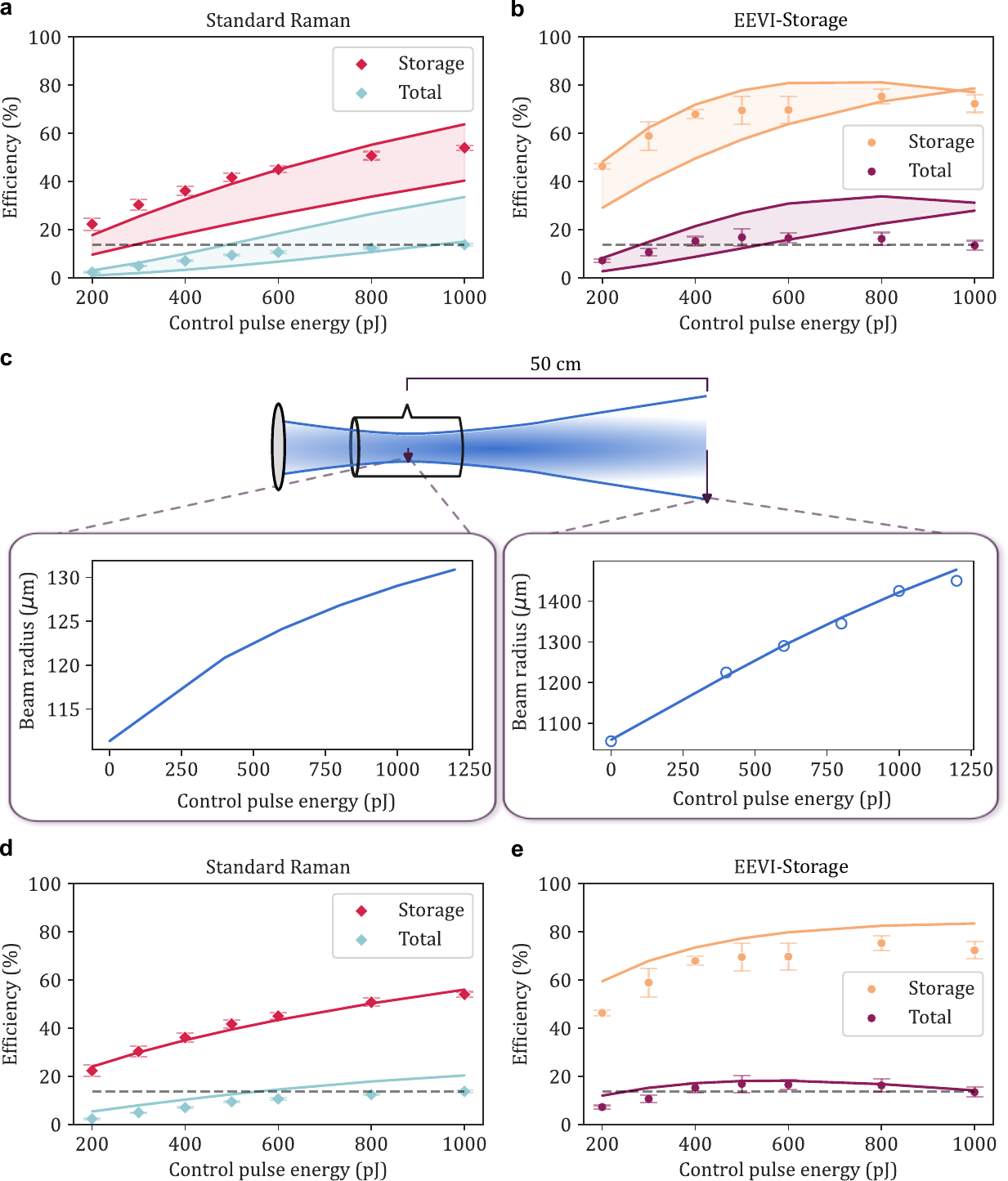}
\caption{Modeling experimental results. \textbf{a} Experimentally measured Storage (red diamonds) and Total (blue diamonds) efficiency vs control pulse energy of the standard Raman memory described in the main text. The solid lines and shaded region show the predicted efficiencies of the numerical model, with the range determined by the control calibration factor used. The dashed gray line indicates the maximum total efficiency, obtained at a control pulse energy of $1000\,$pJ. \textbf{b} Experimentally measured Storage (orange circles) and Total (purple circles) efficiency vs control pulse energy for EEVI applied to the storage interaction of the Raman memory (EEVI-Storage). The solid line and shaded region show the efficiencies predicted from the numerical model. The gray dashed line is again set to the maximum total efficiency attained in the standard Raman protocol. \textbf{c} Schematic of the control field beam focused by a lens into the Cs vapor cell and then propagating $50\,$cm after, where the beam radius is measured (empty circles) as a function of control pulse energy and plotted on the right most plot. The solid blue line shows a fit obtained from numerically solving the non-linear wave propagation equation. The fit is then used to infer the beam radius at the center of the Cs cell as a function of control pulse energy, which is shown on the left most plot. \textbf{d} Experimentally measured Storage (red diamonds) and Total (blue diamonds) efficiency vs control pulse energy of the standard Raman memory, where the solid lines are the predicted efficiencies from the non-linear model. \textbf{e} Experimentally measured Storage (orange circles) and Total (purple circles) efficiency vs control pulse energy for EEVI-Storage, where the solid lines are the predicted efficiencies from the non-linear model.}
\label{figS5}
\end{figure*}

Following the calibration of optical depth, it remains to calibrate the control field Rabi frequency. To do so, we simulate the standard Raman memory using Eqs.~\ref{eq:MB-Raman} for a range of control pulse energies and compare the computed storage and total efficiencies to experimentally measured values, plotted in Fig.~\ref{figS5}\textbf{a} (red and blue diamonds respectively). We include a Lorentzian spectral filter function to emulate the effect of the etalons used to filter the control in the experiment, which were measured to have a combined filter function of $700\,$MHz full-width half maximum. 

Depending on the energy of the control pulses, we find a factor of $1/5.5$ to $1/7.5$ is needed to calibrate the control field (the range can be explained by non-linear interaction of the control field with the Cs atoms, see later), where we attribute $\sim 1/3$ for averaging over the dipole operator and the remaining due to the affect of focusing through the cell, which is not included in the numerical model. We choose $1/5.5$ for calculating the peak power from peak Rabi frequency in Fig.~\ref{figS6}\textbf{c}. The range of storage and total efficiencies predicted by the numerical model are shown by the red and blue shaded regions in Fig.~\ref{figS5}\textbf{a} respectively. The slight negative curvature seen in both the storage and total efficiencies can be attributed to the control field AC Stark shifting the storage state out of two photon resonance. The accumulation of spin wave at the input of the ensemble, which is not optimum for read out in the forwards direction, acts to reduce the total efficiency, but a larger reduction is caused by the retrieved signal being frequency shifted with respect to the central transmission of the etalons as a result of AC Stark shift induced by the control field. Another possible reason for reduction of total efficiency, not accounted for in the numerical model, is the pump field not being adequately switched off during the experiment, leading to reduction of the spin wave. The highest total efficiency in Fig.~\ref{figS5}\textbf{a} is marked by a horizontal dashed gray line.

In Fig.~\ref{figS5}\textbf{b} we plot the corresponding storage efficiency (orange circles) and total efficiency (purple circles) results for EEVI applied to the Raman storage interaction, where again the shaded regions indicate the range in predictions of the numerical model, using the same range in control calibration as the standard Raman memory. Spin wave accumulation at the input to the cell, as well as loss of the frequency shifted retrieved signal due to the collection etalons, also contribute to the reduced total efficiency for EEVI-storage, though the degree of spin wave accumulation is larger due to the majority of interference during the storage interaction being at the input to the cell. For comparison the highest total efficiency of the equivalent standard Raman memory is indicated by a dashed gray line, showing a modest improvement offered by EEVI-Storage, but at much lower control pulse energy.

\paragraph{Non-linear model}
This simple calibration of the control field fails to capture the measured efficiencies at high control field energy, and the range in calibration factors can lead to quite a large error in predicted efficiency. Both of these problems can be explained by non-linear interaction of the control field with the Cs atoms, causing \textit{self-defocusing} and \textit{self-phase modulation}.  Illustrated in Fig.~\ref{figS5}\textbf{c} is a schematic showing the control field being focused by a lens into the Cs vapor cell and propagating $50\,$cm after, where we measured the beam radius as a function of control pulse energy. We plot this in the rightmost plot of  Fig.~\ref{figS5}\textbf{c} (open circles). There is a clear divergence in beam radius as the control pulse energy is increased, caused by self-defocusing. By numerically solving the non-linear wave propagation equation, we can model the beam radius $50\,$cm after the Cs cell, as a function of average control power. The non-linear wave propagation equation, assuming no linear absorption, is given by~\cite{Wu2022},
\begin{equation}
    \frac{\partial \mathcal{E}(x, z)}{\partial z} - \frac{i}{2 k}\frac{\partial^{2} \mathcal{E}(x, z)}{\partial x^{2}} - ik \frac{n_{2}}{n_{0}}|\mathcal{E}(x, z)|^{2}\mathcal{E} = 0 \, ,
\end{equation}
where $\mathcal{E}$ denotes a classical electric field, $x$ indicates the transverse dimension of the beam, $k$ the wavevector, $n_{0}$ the linear refractive index and $n_{2}$ the non-linear refractive index, which is non-zero for $z\in[0, L]$ inside the Cs cell. In this model, we have ignored the time dynamics of the control pulses and atom response, instead assuming a constant control field intensity and steady state of the atoms.  We account for saturation of $n_{2}$ through the relation,
\begin{equation}
    n_{2}(I) = 
    \begin{cases}
    \frac{n_{2}(0)}{1+\frac{I}{I^{\Delta}_{\rm{sat}}}} & z\in[0, L]\\
    0 & \rm{otherwise} \, ,
    \end{cases}
\end{equation}
with $I$ denoting the intensity of the control field using the standard beam waist, and $I^{\Delta}_{\rm{sat}} = I_{\rm{sat}}\left(1+4 \pi^{2} \left(\frac{\Delta}{\delta}\right)^{2} \right)$~\cite{Wu2022}, for a saturation intensity $I_{\rm{sat}} = 1.1\,$mW\,cm$^{-2}$~\cite{Steck2024}, detuning $\Delta = 9.2\,$GHz and Cs ensemble Doppler width $\delta = 290\,$MHz. The resulting beam radius from this model is shown in Fig.~\ref{figS5}\textbf{c} (solid blue), giving an estimated $n_{2}(0)=-7.5 \times 10^{-8}\,$cm$^{2}\,$W$^{-1}$ which is consistent with experimentally measured values~\cite{Ara2013}. This fit was then used to infer the radius at the center of the cell, which is plotted in the left plot of  Fig.~\ref{figS5}\textbf{c}. We observe a $16\%$ change in the beam waist at the focus, which together with self-phase modulation, explains the observed reduction in the control field calibration value with increasing energy in Fig.~\ref{figS5}(\textbf{a},\textbf{b}).\\

We now incorporate the non-linear effects into the numerical model by modifying the control pulse intensity according to the inferred radius from our steady state non-linear model, together with accounting for self-phase modulation given by,
\begin{equation}
    \Delta(t) = \Delta + \frac{4\pi L n_{2} I_{0}}{\lambda_{0} \sigma^{2}}\, t\, \exp\left(\frac{-t^{2}}{\sigma^{2}}\right) \, ,
\end{equation}
with $I_{0}$ the peak intensity of the Gaussian control pulse, $\lambda_{0}$ the vacuum wavelength and $\sigma$ defines the $1/e$ amplitude width. We obtain a much better match between the numerical model and the experimental results, shown in  Fig.~\ref{figS5}\textbf{d} and  Fig.~\ref{figS5}\textbf{e}. While this model explains the observed behavior for standard Raman and EEVI-Storage, due to not accounting for the time dynamics of the non-linear effect, the model does not generalize well to experimental results performed with differing control pulse amplitudes. 

\subsection*{Optimization}

\paragraph{Krotov method}
In Fig.~6 of the main text, we presented the numerical results of optimizing the control pulse shapes using a local gradient ascent optimizer. The local optimizer we employ is the Krotov method, explained in more detail in Refs.~\cite{Gorshkov2008, Rakher2013}, but summarized briefly here. We will omit any operator notation for readability.

We seek to maximize the total efficiency $\eta$, defined as,
\begin{equation}
    \eta = \int_{\tau\in R} d\tau \, \hat{\mathcal{E}}^{*}(z=L, \tau)\hat{\mathcal{E}}(z=L, \tau) \,
\end{equation}
in which $L$ is the length of the Cs ensemble, the integral is over the time interval of the final read process (this will be different depending on the precise protocol being optimized). We must optimize $\eta$ while ensuring dynamic variables are constrained according to the equations of motion. The full objective function may be written as,
\begin{equation}
J = \eta + f_{W_{1}} + f_{W_{2}} + f_{R_{1}} + f_{R_{2}} \, ,
\end{equation}
where the functional $f_{X}$ denotes the Lagrangian multipliers and corresponding equations of motion over the time interval $X$, for example,
\begin{equation}
    \begin{split}
        f_{W_{1}} =
        &\sum_{v} \int_{0}^{L}dz\int_{\tau\in W_{1}}d\tau \,\Bar{S}^{*}\left[-\partial_{\tau}{S} - i \frac{\sqrt{d}}{\Gamma}\Omega_{W_{1}}^{*} \mathcal{E} - \frac{|\Omega_{W_{1}}|^{2}}{\Gamma} S\right] + c.c.\\
        &+ \int_{0}^{L}dz\int_{\tau\in W_{1}}d\tau \, \Bar{\mathcal{E}}^{*}\left[-\partial_{z}\mathcal{E}+\sum_{v}\left(i \frac{\sqrt{d}}{\Gamma}\Omega_{W_{1}} S - \frac{d}{\Gamma}\mathcal{E}\right)\right] + c.c. \, ,
    \end{split}
\end{equation}
where $\Bar{\mathcal{E}}$ and $\Bar{S}$ denote the Lagrangian multipliers, $c.c.$ refers to the complex conjugate and we have removed any explicit $z, \tau$ and $v$ from the variables for readability. Here $\Omega_{W_{1}}$ is the complex control field defined on the time interval $W_{1}$. The equations of motion for the Lagrangian multipliers can be shown to be the time and space reversed counterparts of the standard equations of motion, where the input signal boundary condition is the time-reversed output of the forward propagating simulation.\\

From the objective functional $J$, we may derive an analytic expression for the gradient on each of the time intervals, which has the general form,
\begin{equation}
    \frac{\partial J}{\partial \Omega_{X}^{*}(\tau)} = \sum_{v} -i 
\frac{|\Omega_{W_{1}}|^{2}}{\Gamma}    \int_{0}^{1}dz\left( \Bar{\mathcal{E}}_{X} \frac{\sqrt{d}}{\Gamma^{*}} S^{*}_{X} - \Bar{S}^{*}_{X} \frac{\sqrt{d}}{\Gamma} \mathcal{E}_{X} - 2\,\textrm{Re}\left[\Bar{S}^{*}_{X} \frac{\Omega_{X}}{\Gamma} S_{X} \right] \right) \, ,
\end{equation}
where the subscript $X$ is used to indicate the time interval which the associated variable is defined on. The control field may then be iteratively improved using the replacement rule,
\begin{equation}
    \Omega^{(j+1)}_{X}(\tau) \leftarrow \Omega^{(j)}_{X}(\tau) + \alpha \frac{\delta J}{\delta \Omega^{(j)}_{X}(\tau)} \, ,
\end{equation}
for a step size $\alpha$. We dynamically change $\alpha$ according to the Wolfe conditions~\cite{Rakher2013} and terminate the optimization when the total efficiency over five iterations has changed by less than $0.01\%$.

In practice, we simulate the system forwards in time and space with an initial guess of the control fields, after which the time-reversed output is used as the initial condition for the solving the system backwards in time and space. The two simulations can then be used to construct a gradient for the control field defined on each time interval, which is used to update each of the control fields. For the standard Raman protocol, the retrieval pulse is fixed as a sigmoid function with a Rabi frequency of $16\,$GHz as we are only concerned with the integrated output signal intensity and not the output temporal mode. However for EEVI-Raman, the signal temporal mode is important at each stage of the protocol, so we optimize all four control fields. For Fig.~6\textbf{c} in the main text, to limit the maximum available Rabi frequency, any point of the updated control field that is above the maximum Rabi frequency is reduced to the maximum before the next iteration is performed.

\begin{figure*}[t]
\centering
\includegraphics[width=0.8\textwidth]{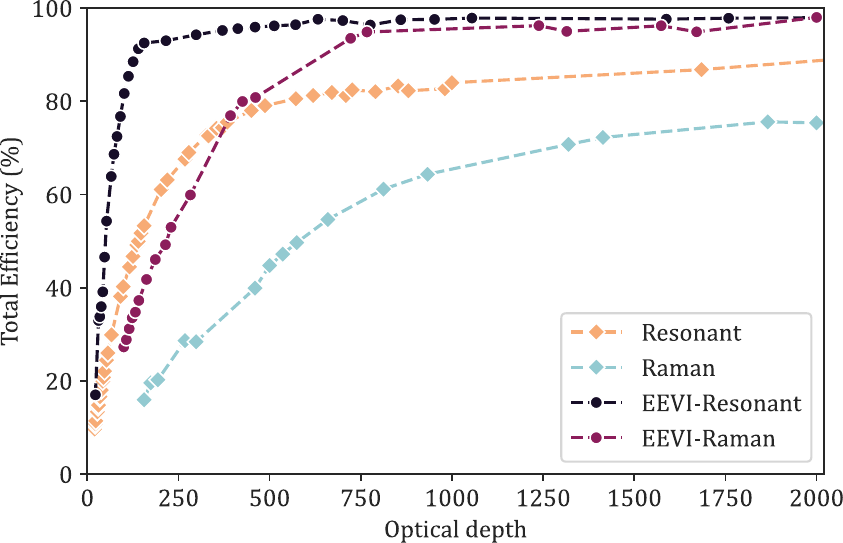}
\caption{Optimized total efficiency for forwards retrieval as a function of optical depth for a standard resonant memory (orange diamonds), Raman memory (blue diamonds), EEVI-Resonant (black circles) and EEVI-Raman (blue circles) assuming $100\%$ loop transmission.}
\label{figS6}
\end{figure*}

This optimization procedure is quite general, and can be applied to memory protocols in which the signal and control are near resonance, see Refs.~\cite{Gorshkov2008, Rakher2013}, as is the case in memories based off Electromagnetically-Induced-Transparency (EIT) and Autler-Townes-Splitting. To investigate how EEVI may be used to improve near-resonant memories, we apply the same gradient ascent procedure for maximizing total efficiency in the forwards direction, but with the signal and control on-resonance. We keep the input signal at $1\,$GHz bandwidth. The results for a standard resonant memory are shown in Fig.~\ref{figS6} (orange diamonds). Due to the large bandwidth and low optical depth, the memory process is predominantly non-adiabatic, and so the optimized pulses resemble the ATS memory protocol. However at large optical depths the memory becomes more adiabatic and the dynamics begin to resemble EIT quantum memories. Even for low optical depth where the memory is very non-adiabatic, we see applying EEVI to both the storage and retrieval memory processes (black circles) leads to a drastic increase in total efficiency. For an optical depth of 100, EEVI-Resonant shows an increase from $40\%$ to $80\%$. Interestingly, while at low optical depth the standard resonant memory performs better than EEVI-Raman, above an optical depth of 400, EEVI-Raman quickly surpasses resonant memories.  

It is important to note that this optimization was for a simple three-level model. While for Raman memories this is a reasonable approximation, for resonant memories multiple closely spaced excited states would result in more complex dynamics, particularly with the large signal bandwidth used here. However, for lower bandwidth signals or with the use of strong magnetic fields, where a single excited state can be isolated, EEVI may offer significantly higher efficiencies at low optical depth.

\paragraph{Modal capacity}
In Fig.~6\textbf{d} of the main text, we examine the modal capacity of the memory. For the standard Raman memory, we optimize the shape of the control write pulse for storing a Gaussian input signal, while fixing the control pulse energy~\cite{Gorshkov2008}. For each control pulse energy (or equivalently storage efficiency) we construct the storage kernel using the first 10 Hermite-Gaussian temporal modes as our input basis, and calculate the resulting spin waves. Through single-value decomposition, we can deduce the eigenfunctions of the storage kernel and corresponding eigenvalues. The square of the $i$'th eigenvalue gives the storage efficiency of the $i$'th eigenfunction, therefore the highest possible storage efficiency is given by the first eigenvalue squared. The Schmidt number $S$ is calculated using the expression:
\begin{equation}
    S =  \frac{1}{\sum_{i} \eta_{i}^{2}} \, ,
\end{equation}
where $\eta_{i}$ corresponds to the storage efficiency of the $i$'th eigenmode after normalizing $\sum_{i} \eta_{i} = 1$~\cite{NunnThesis}. The same procedure is repeated for EEVI-Raman, where both write pulses are fixed to the same energy and the shapes optimized. The storage kernel is constructed for the entire write process with the signal and spin wave interference set to maximize the storage efficiency.

\bibliography{bibliography}

\end{document}